\documentclass[structabstract]{aa}

\usepackage{graphicx}
\usepackage{txfonts}
\usepackage{natbib}
\bibpunct{(}{)}{;}{a}{}{,}

\usepackage{mathbbol}

\begin{document}

\title{Radiative transfer with scattering for domain-decomposed 3D MHD simulations of cool stellar atmospheres}
\subtitle{Numerical methods and application to the quiet, non-magnetic, surface of a solar-type star}

\author{
W. Hayek\inst{1,2,3}
\and M. Asplund\inst{1}
\and M. Carlsson\inst{3}
\and R. Trampedach\inst{4,2}
\and R. Collet\inst{1}
\and B. V. Gudiksen\inst{3}
\and V. H. Hansteen\inst{3}
\and J. Leenaarts\inst{5}
}
\institute{
Max Planck Institut f\"ur Astrophysik, Karl-Schwarzschild-Str. 1, D-85741 Garching, Germany\\
\email{[hayek,asplund,remo]@mpa-garching.mpg.de}
\and
Research School of Astronomy \& Astrophysics, Cotter Road, Weston Creek 2611, Australia\\
\and
Institute of Theoretical Astrophysics, University of Oslo, P.O. Box 1029 Blindern, N-0315 Oslo, Norway\\
\email{[mats.carlsson,b.v.gudiksen,viggo.hansteen]@astro.uio.no}
\and
JILA, University of Colorado, 440 UCB, Boulder, CO 80309-0440, U.S.A.\\
\email{trampeda@lcd.colorado.edu}
\and
Astronomical Institute, Utrecht University, Postbus 80 000, NLÐ3508 TA Utrecht, The Netherlands\\
\email{j.leenaarts@uu.nl}
}

\date{\today}

\abstract
{}
{We present the implementation of a radiative transfer solver with coherent scattering in the new \texttt{BIFROST} code for radiative magneto-hydrodynamical (MHD) simulations of stellar surface convection. The code is fully parallelized using MPI domain decomposition, which allows for large grid sizes and improved resolution of hydrodynamical structures. We apply the code to simulate the surface granulation in a solar-type star, ignoring magnetic fields, and investigate the importance of coherent scattering for the atmospheric structure.}
{A scattering term is added to the radiative transfer equation, requiring an iterative computation of the radiation field. We use a short-characteristics-based Gauss-Seidel acceleration scheme to compute radiative flux divergences for the energy equation. The effects of coherent scattering are tested by comparing the temperature stratification of three 3D time-dependent hydrodynamical atmosphere models of a solar-type star: without scattering, with continuum scattering only, and with both continuum and line scattering.}
{We show that continuum scattering does not have a significant impact on the photospheric temperature structure for a star like the Sun. Including scattering in line-blanketing, however, leads to a decrease of temperatures by about 350\,K below $\log_{10}\tau_{5000}\lesssim-4$. The effect is opposite to that of 1D hydrostatic models in radiative equilibrium, where scattering reduces the cooling effect of strong LTE lines in the higher layers of the photosphere. Coherent line scattering also changes the temperature distribution in the high atmosphere, where we observe stronger fluctuations compared to a treatment of lines as true absorbers.}
{}

\keywords{Radiative transfer -- Stars: atmospheres -- Sun: atmosphere}

\maketitle

\section{Introduction}

The atmospheres of late-type stars form the transition from the opaque convective envelope to the interstellar medium. Hot rising plasma transports heat to the surface, becomes transparent and looses its entropy through radiative cooling. Gravity accelerates the cooled gas back into the star, carrying kinetic energy inward and forcing the convective flow. By taking over heat transport and removing entropy, the radiation field therefore indirectly drives convection \citep{Steinetal:1998}, making radiative and hydrodynamical processes equally important at the surface. Magnetic fields have strong impact on the higher atmosphere and cause local phenomena in the surface granulation, such as spots and pores.

The classical numerical models of cool stellar atmospheres in 1D focused on a detailed description of radiative transfer, with two prominent examples being the \texttt{MARCS} code \citep{Gustafssonetal:1975} and the \texttt{ATLAS} code \citep{Kurucz:1979}. Assuming a plane-parallel or spherical-symmetric stratification, they include only a rudimentary treatment of convective energy transport in cool stellar atmospheres. Subsequent updates of these models \citep[e.g.,][]{Kurucz:1996,Gustafssonetal:2008} benefit from the largely increased computational power, refining the treatment of the strongly wavelength-dependent line opacities. Newer codes, such as \texttt{PHOENIX} \citep{Hauschildtetal:1999} can also include departures from local thermodynamic equilibrium (LTE) in the radiative transfer computation and the absorber populations. The 1D models have not only provided growing insight into the physical environment at the surface of cool stars, but have also become a standard tool for chemical abundance analyses. The wide variety of applications includes studies of galactic chemical evolution and of the origin of the elements.

The advent of fully dynamic 3D surface convection simulations has enabled a much more realistic treatment of the hydrodynamical plasma flow, deepening our understanding of convection and eliminating the need for microturbulent and macroturbulent broadening in line formation computations \citep[see, e.g.,][]{Nordlundetal:2009}. The 3D models are capable of accurately reproducing the surface structure of the observed solar granulation with their strongly inhomogeneous surface intensities \citep{Steinetal:1998}. The velocity fields predicted by the 3D simulations lead to a close match with both the observed spectral line bisectors and the broadening of their profiles in the atmospheres of different stars \citep[e.g.,][]{Dravinsetal:1990,Asplundetal:2000,AllendePrietoetal:2002,Ramirezetal:2009}. Recently, impressive agreement between a new synthetic 3D model and solar observations has been found in a detailed comparison of spectral line shifts, equivalent widths and center-to-limb variations for normalized line profiles \citep{Pereiraetal:2009b,Pereiraetal:2009a}. In essentially all cases, this 3D model reproduced the observations with an accuracy that is comparable to the semi-empirical model of \citet{Holwegeretal:1974}, which is traditionally used in spectroscopy of the solar photosphere.

The accuracy of the treatment of radiation in 3D, however, is still strongly limited by the available computational power. Radiative transfer easily becomes the most computationally expensive part of a simulation, since the equations must be solved for a considerably larger set of transport directions compared to hydrodynamics, and non-grey opacities must be accounted for in realistic simulations. Most of the currently existing 3D radiative (M)HD codes therefore assume LTE and capture the atmospheric height dependence of continuum and line opacities using the opacity binning method \citep[e.g.,][]{Nordlund:1982,Ludwig:1992}: the problem of computing the monochromatic radiation field for a larger number of wavelengths is reduced to the numerical solution of the radiative transfer equation for typically 5 opacity bins. \citet{Skartlien:2000} extended the opacity binning method to include coherent scattering, and showed its importance in the solar chromosphere using a 3D radiative transfer solver for parallel shared-memory architectures.

Modern large-scale computer clusters use distributed memory architectures to handle the growing complexity of scientific simulations, allowing, e.g., self-consistent MHD models of the solar chromosphere, transition region and corona \citep{Hansteen:2004,Hansteenetal:2007} or detailed hydrodynamical models of giant stars \citep{Colletetal:2007}. We present a new fully MPI-parallelized radiative transfer solver with coherent scattering for the new \texttt{BIFROST} code for time-dependent 3D MHD simulations of cool stellar atmospheres (Gudiksen et al., in preparation).

In Sect.~\ref{sec:rt} and \ref{sec:implementation}, we discuss the physics of the radiative transfer model and its implementation in the MHD code. Section~\ref{sec:opacity} describes the most important continuous and line opacity sources that we include in our simulations. Section~\ref{sec:sun} describes the application of the \texttt{BIFROST} code to model the atmosphere of a solar-type star using radiative transfer calculations with scattering, and discusses the effects on the temperature structure.

\section{Radiative transfer with scattering and the radiative flux divergence}\label{sec:rt}

\subsection{The radiative transfer equation}\label{sec:radtrans}

Hydrodynamical simulations of cool stellar atmospheres need to cover several pressure scale heights above and below the optical surface to minimize the effect of the boundaries on the granulation flow. The exponential density stratification causes the optical depth of the plasma to span about 15 orders of magnitude from the highest to the lowest layers of the simulation. The radiative transfer problem must therefore be solved in very different physical environments: in the extremely optically thick diffusion region at the bottom of the simulation box, all photons are thermalized. At the top, the atmosphere is mostly optically thin and mainly photons in the strongest lines interact with the gas. For the bulk of the photons, the transition between these two domains is rapid; it is confined to a thin layer which appears corrugated due to the different geometrical depth variation of opacities in upflows and downflows \citep{Steinetal:1998}.

Radiative transfer is, in general, a time-dependent process, which needs to be treated simultaneously with the hydrodynamics. However, the timescale of photon propagation over a mean free path length, $t_{\lambda}=(c\chi_{\lambda})^{-1}$, where $\chi_{\lambda}$ is the monochromatic opacity and $c$ is the speed of light, is orders of magnitude shorter than any hydrodynamical timescale. Radiative transfer therefore decouples from the hydrodynamics and is well approximated by a time-independent problem, described by a radiative transfer equation for the monochromatic specific intensity $I_\lambda(\vec{x},\vec{\hat{n}})$ in direction $\vec{\hat{n}}$:
\begin{equation}
\vec{\hat{n}\cdot\nabla}I_{\lambda}(\vec{x},\vec{\hat{n}})=-\chi_{\lambda}(\vec{x})I_{\lambda}(\vec{x},\vec{\hat{n}})+j_{\lambda}(\vec{x},\vec{\hat{n}}),
\label{eqn:RT}
\end{equation}
where $j_{\lambda}$ denotes the local emission at wavelength $\lambda$ \citep[see, e.g.,][]{Mihalas:1978,Rutten:2003}. The left-hand side of Eq.~(\ref{eqn:RT}) is defined in the rest frame of the model atmosphere. The source and sink terms of the right-hand side are naturally described in the co-moving frame of the flowing gas. The consequent Doppler shifts are difficult to treat in 3D time-dependent simulations due to restrictions in computational power, requiring us to compute wavelength-integrated quantities in the opacity binning approximation (see below). We therefore assume a static medium, neglecting possible influences of the velocity field.

The extinction of photons is described, as customary, through the absorption coefficient $\kappa_{\lambda}$ and the scattering coefficient $\sigma_{\lambda}$, which combine to the gas opacity,
\begin{equation}
\chi_{\lambda}(\vec{x})=\kappa_{\lambda}(\vec{x})+\sigma_{\lambda}(\vec{x}),
\end{equation}
and give rise to the definition of the photon destruction probability
\begin{equation}
\epsilon_{\lambda}(\vec{x})=\frac{\kappa_{\lambda}(\vec{x})}{\kappa_{\lambda}(\vec{x})+\sigma_{\lambda}(\vec{x})}.
\end{equation}
Recasting the optical path $ds=\vec{\hat{n}\cdot dx}$ along a ray in direction $\vec{\hat{n}}$ into the optical depth $d\tau_{\lambda}=\chi_{\lambda}ds$ along that direction, gives Eq.~(\ref{eqn:RT}) the form
\begin{equation}
\frac{dI_{\lambda}}{d\tau_{\lambda}}(\tau_{\lambda})=-I_{\lambda}(\tau_{\lambda})+S_{\lambda}(\tau_{\lambda}),
\label{eqn:RTtau}
\end{equation}
with the source function $S_{\lambda}\equiv j_{\lambda}/\chi_{\lambda}$. For the numerical computation, we employ the formal solution
\begin{equation}
I_{\lambda}(\tau_{\lambda})=I_{\lambda}(\tau_{\mathrm{u},\lambda})e^{-(\tau_{\lambda}-\tau_{\mathrm{u},\lambda})}+\int_{\tau_{\mathrm{u},\lambda}}^{\tau_{\lambda}}S_{\lambda}(t)e^{t-\tau_{\lambda}}dt,
\label{eqn:fs}
\end{equation}
where $I_{\lambda}(\tau_{\mathrm{u},\lambda})$ is the incident intensity at the upwind end of the ray at optical depth $\tau_{\mathrm{u},\lambda}<\tau_{\lambda}$.

The source function $S_{\lambda}$ at optical depth $\tau_{\lambda}$ in direction $\vec{\hat{n}}$ includes local thermal radiation from the gas and coherent scattering of photons:
\begin{equation}
S_{\lambda}=\frac{\sigma_{\lambda}}{4\pi\chi_{\lambda}}\int_{S^{2}}\phi(\vec{\hat{n}},\vec{\hat{n}'})I_{\lambda,\vec{\hat{n}'}}d\Omega'+\frac{\kappa_{\lambda}}{\chi_{\lambda}}B_{\lambda}=\left(1-\epsilon_{\lambda}\right)J_{\lambda}+\epsilon_{\lambda}B_{\lambda},
\label{eqn:sourceterm}
\end{equation}
where scattered radiation from direction $\vec{\hat{n}'}$ contributes with weight $\phi$ in the integral over the unit sphere $S^2$, $B_{\lambda}$ denotes the Planck function, and $J_{\lambda}$ is the mean intensity. The second equality holds for isotropic angular redistribution of radiation ($\phi=1$). For $\epsilon_{\lambda}<1$, the source function depends on $J_{\lambda}$ and thus, through the non-locality of radiative transfer, on radiation processes in the entire simulation domain. This turns Eq.~(\ref{eqn:RTtau}) from an ordinary differential equation into an integro-differential equation.

Current limitations of available computing resources require the assumption of isotropic coherent scattering. Continuum processes in cool stellar atmospheres and very strong lines fulfill this restriction in very good or reasonable approximation, respectively, due to their weak wavelength dependence. Intermediate and weak lines are more accurately treated in complete spectral redistribution.


\subsection{The radiative flux divergence and the wavelength integral}

Absorption and thermal emission of radiation couples the stellar plasma with the radiation field through the transfer of heat. Photon energies in cool stars are too small to exert a significant force on the fluid compared to the gas pressure and gravity; the coupling is therefore sufficiently described by adding a radiative heating term $Q_{\mathrm{rad}}$ to the energy equation.

Evaluating the first moment of Eq.~(\ref{eqn:RT}) and using the above definitions yields
\begin{equation}
-\vec{\nabla\cdot F}_{\lambda}=4\pi\chi_{\lambda}\left(J_{\lambda}-S_{\lambda}\right)=4\pi\epsilon_{\lambda}\chi_{\lambda}\left(J_{\lambda}-B_{\lambda}\right),
\label{eqn:divF}
\end{equation}
where $\vec{\nabla\cdot F}_{\lambda}$ is the local monochromatic radiative flux divergence. The second equality holds in the case of the coherent scattering source function (Eq.~(\ref{eqn:sourceterm})). The scattering term does not contribute to heat exchange by definition, reducing radiative heating and cooling by a factor of $\epsilon_{\lambda}$ compared to the case where $S_{\lambda}=B_{\lambda}$.

Integrating the monochromatic flux divergence in Eq.~(\ref{eqn:divF}) over the whole wavelength spectrum of the star yields the local heating rate $Q_{\mathrm{rad}}$:
\begin{equation}
Q_{\mathrm{rad}}=-\int_{0}^{\infty}\vec{\nabla\cdot F}_{\lambda}d\lambda.
\label{eqn:wavint}
\end{equation}
In the optically thick regime, where radiative transfer is diffusive, this integral may be simplified with good accuracy by assuming the Rosseland mean opacity in the so-called gray approximation. However, gray opacities are not sufficient for a realistic treatment of the height-dependent line-blanketing above the surface, where the atmospheric structure is very sensitive to the radiation field. Atomic and molecular lines are important opacity sources in this region, changing the radiative heating and cooling compared to the simplified case of a gray atmosphere \citep[see, e.g.,][for a detailed discussion]{Voegleretal:2004}. The current version of the \texttt{MARCS} 1D atmosphere code uses the opacity sampling technique \citep{Peytremann:1974}, which approximates the spectrum through statistical sampling at $\sim100\,000$ wavelength points. This resolution is sufficient to capture continuum absorption and line-blanketing without bias, at least in the lower parts of the atmosphere where the spectral distribution of absorbers is sufficiently widespread. Stellar atmosphere models in 3D do not allow for such a detailed treatment yet, since a single formal solution is many orders of magnitude more expensive to compute: the radiative transfer equation in our radiation-hydrodynamical model (Sect.~\ref{sec:hydromodel}) is solved for $240\times240$ columns and 24 transport directions, which is equivalent to $\sim10^{5}$ 1D calculations for each time step.

\citet{Nordlund:1982}, \citet{Ludwig:1992} and \citet{Skartlien:2000} have described opacity binning techniques, where wavelength integration is performed over subsets of the spectral range before the solution of the radiative transfer equation, and the radiation field is computed for only a few mean opacities  instead of the full spectrum. We will give a brief description of the technique in the following; see \citet{Skartlien:2000} for a more detailed discussion.

Integrating the radiative transfer equation (Eq.~(\ref{eqn:RT})) over wavelength leads to the definition of a mean opacity, mean scattering coefficient and mean absorption coefficient:
\begin{eqnarray}
\chi^{\mathrm{I}}&=&\frac{\int\chi_{\lambda}I_{\lambda}d\lambda}{\int I_{\lambda}d\lambda}\label{eqn:intmean}\\
\sigma^{\mathrm{J}}&=&\frac{\int\sigma_{\lambda}J_{\lambda}d\lambda}{\int J_{\lambda}d\lambda}\\
\kappa^{\mathrm{B}}&=&\frac{\int\kappa_{\lambda}B_{\lambda}d\lambda}{\int B_{\lambda}d\lambda}.
\end{eqnarray}
The intensity-weighted mean opacity $\chi^{\mathrm{I}}$ and the mean-intensity-weighted mean scattering coefficient $\sigma^{\mathrm{J}}$ depend on the unknown radiation field $I_{\lambda}$ and its angular average $J_{\lambda}$, which must be estimated: we use 1D radiative transfer calculations on the mean stratification of the atmosphere (see Sect.~\ref{sec:1Dmean}), which yield approximations for $\chi^{\mathrm{I}}\approx\chi^{\mathrm{J,1D}}$ and $\sigma^{\mathrm{J}}\approx\sigma^{\mathrm{J,1D}}$.

These three mean coefficients represent absorption, scattering and thermal emission of photons with good accuracy where the stellar atmosphere is optically thin across the spectrum. However, $\chi^{\mathrm{J,1D}}$ does not ensure a correct total radiative energy flux at optical depths $\tau\gg1$ where radiative transfer is diffusive. It needs to be replaced by the Rosseland mean opacity, defined as the weighted harmonic mean
\begin{equation}
\chi^{\mathrm{R}}=\frac{\int\left(dB_{\lambda}/ds\right)d\lambda}{\int\left(1/\chi_{\lambda}\right)\left(dB_{\lambda}/ds\right)d\lambda}.
\label{eqn:rossmean}
\end{equation}
We consequently use a $\tau$-weighted sum of the two quantities $\chi^{\mathrm{J,1D}}$ and $\chi^{\mathrm{R}}$. The geometrical depth of the transition between the two regimes near $\tau\approx1$ varies quickly with wavelength where spectral lines are present, and it is not sufficient to consider only a single pair of mean opacities $\chi^{\mathrm{J,1D}}$ and $\chi^{\mathrm{R}}$. The opacity binning method therefore defines several opacity groups, where each member reaches unit optical depth ($\tau_{\lambda}=1$) at a similar geometrical depth. The integrals in Eq.~(\ref{eqn:intmean}) - Eq.~(\ref{eqn:rossmean}) are then evaluated only for a set of member wavelengths $\{\lambda_{i}\}$ in each bin $i$, which does not have to be continuous.

Depending on the height range of the stellar atmosphere model and the wavelength selection method, it turns out that about 5 such opacity bins are enough to capture the essence of the line-blanketing and continuum opacity and to obtain a realistic temperature structure \citep{Voegleretal:2004}. More recent atmosphere models have been extended to 12 bins \citep{Caffauetal:2008}. For the simulations presented in this work, we compute radiative transfer with 12 bins, where wavelengths are sorted not only by the geometrical height of the monochromatic optical surface, but also by wavelength, separating opacities in the UV, visual and infrared bands (Trampedach et al., in preparation).

It is difficult to assess the quality of the opacity binning method in realistic 3D simulations: deviations of the resulting radiative heating rates $Q_{\mathrm{rad}}$ from an accurate monochromatic solution have a height-dependent impact on the temperature structure (see Sect.~(\ref{sec:sun})), making the long-term behavior of the simulation hard to predict. The agreement of 3D model atmospheres with various observational tests indicates that opacity binning still yields a reasonable estimate for the line-blanketing.


\section{The numerical implementation}\label{sec:implementation}

The large variety of radiative transfer models for astrophysical problems inspired the development of very different analytical and numerical methods to obtain the radiation field (see, e.g., \citet{Wehrseetal:2006} for an overview). For our given problem of computing radiative heating rates as the flux divergence $-\vec{\nabla\cdot F}$ of a time-independent radiation field in 3D, the direct solution of Eq.~(\ref{eqn:RTtau}) yields accurate results with efficient numerical schemes.

Characteristics methods, which solve the transfer problem along a discrete set of light rays to capture the anisotropy of the radiation field in the optically thin atmosphere, are a popular choice in stellar atmosphere models. \citet{Nordlund:1982} and \citet{Skartlien:2000} use Feautrier-type differential radiative transfer solvers \citep{Feautrier:1964} for solving Eq.~(\ref{eqn:RTtau}) on long characteristics. They span across the entire simulation domain, which is an obstacle for a domain-decomposed parallelization of the MHD code (see Sect.~\ref{sec:gsmpi} below). \citet{Brulsetal:1999}, \citet{Voegleretal:2005} and \citet{Muthsametal:2010} employ the short characteristics method \citep{Mihalasetal:1978,Olsonetal:1987,Kunaszetal:1988}, where the radiative transfer equation is solved on characteristics which only extend to the adjacent upwind and downwind grid layers. This method is required by our choice of iteration technique for an efficient solution of the scattering problem.


\subsection{Short characteristics}\label{sec:SC}

The short characteristics method employs the formal solution (Eq.~(\ref{eqn:fs})) of the monochromatic radiative transfer equation (Eq.~(\ref{eqn:RTtau})) to compute the radiation field at the center of a three-point ray for a known source function $S_{\lambda}$. The discretization is performed by interpolating the source function for a given wavelength $\lambda$ (or bin number) along the ray using a second-order B\'ezier curve \citep[see, e.g., the discussion in][]{Auer:2003}
\begin{equation}
S(t)=(1-t)^2S_{\mathrm{u}}+t^2S_{0}+2t(1-t)S_{\mathrm{c}},
\label{eqn:Bezier}
\end{equation}
where $S_{\mathrm{u}}$ and $S_{0}$ are the upwind and local source functions and $t=(\tau-\tau_{\mathrm{u}})/(\tau_{0}-\tau_{\mathrm{u}})$ is the curve parameter. $S_{\mathrm{c}}$ is a control point, which shapes the interpolating curve by restricting it to the convex hull laid out by $S_{\mathrm{u}}$, $S_{\mathrm{c}}$ and $S_{0}$. This characteristic of B\'ezier curves may be exploited to detect and suppress overshoots, which destabilize the numerical solution at places in the atmosphere where strong opacity and temperature gradients occur. Inserting Eq.~(\ref{eqn:Bezier}) into the formal solution (Eq.~(\ref{eqn:fs})), evaluating the integral and reordering the terms yields the discretized expression
\begin{equation}
I(\tau)=I(\tau_{\mathrm{u}})e^{-(\tau-\tau_{\mathrm{u}})}+\Psi_{\mathrm{u}}S_{\mathrm{u}}+\Psi_{0}S_{0}+\Psi_{\mathrm{d}}S_{\mathrm{d}}.
\label{eqn:discretefs}
\end{equation}
The shape of the three interpolation coefficients $\Psi_{\mathrm{u}}$, $\Psi_{0}$ and $\Psi_{\mathrm{d}}$ for the upwind, center and downwind source functions depends on the control point $S_{\mathrm{c}}$. Choosing
\begin{equation}
S_{\mathrm{c}}=S_{0}-\frac{\tau_{0}-\tau_{\mathrm{u}}}{2}S_{0}',
\end{equation}
where $S_{0}'$ is the centered derivative on the three-point stencil $(S_{\mathrm{u}},S_{0},S_{\mathrm{d}})$, yields second-order interpolation. It is used where no overshoots happen and correctly reproduces the diffusion approximation at optical depths $\tau\gtrsim30$ (see Appendix~\ref{sec:scconst} for the detailed shape of the $\Psi$ coefficients). In the optically thin atmosphere where $\tau\lesssim10^{-3}$, a second-order expansion of the $(1-e^{-\tau})$ terms in the $\Psi$ constants stabilizes the solver, which may therefore be implemented with single precision floating point numerics throughout the simulation domain. Optical depths $\Delta\tau$ along the characteristics are similarly computed using the B\'ezier interpolation technique.

The mean intensities $J$ and the components of the flux vector $\vec{F}$ are computed by approximating the zeroth and first moment integrals by a quadrature sum over selected ray angles (``method of discrete ordinates''),
\begin{equation}
J\approx\frac{1}{4\pi}\sum_{i}w_{i}I(\vec{\hat{n_{i}}}); \hspace{0.5cm} F_{j}\approx\sum_{i}w_{i}I(\vec{\hat{n_{i}}})(\vec{\hat{n_{i}}\cdot\hat{n}_{j}}),
\end{equation}
where $w_{i}$ is the weight of direction $\vec{\hat{n_{i}}}$. The best choice of quadrature depends on the expected anisotropy of the radiation field and on the quantity that needs to be computed. In our case, the components of the flux vector $\vec{F}$ need to be calculated explicitly, requiring the quadrature be invariant to rotation by $\pi/2$ to avoid directional bias. Carlson's A4 quadrature \citep{Carlson:1963} with 3 rays per octant is an appropriate choice and represents the anisotropy with good accuracy.

Short characteristics require knowledge of the upwind intensities $I(\tau_{\mathrm{u}})$ for each ray direction $\vec{\hat{n}}$, on which the sweep direction for a formal solution therefore depends. Interpolation yields all such quantities (Sect.~\ref{sec:intrefine}). Shallow rays, that fail to hit the upwind layer within the grid cells, need to be extended and may cross several cells, possibly across subdomain boundaries. For the first formal solution of a simulation run, a Feautrier-type long characteristics solver delivers boundary intensity estimates; intensities from the previous iteration in the neighbor subdomains are used for all subsequent computations. Once $I(\tau)$ is known along two edges of the current layer, the remaining unknown intensities may be computed away from the boundary through vertical interpolation between the upwind layer and the current layer. It is worth noting that some long characteristics codes turn transport directions around the vertical axis with every time step to avoid numerical artefacts stemming from a fixed set of discrete ordinates. Such an effect is not observed in our short characteristics implementation. Moreover, the anisotropy of the radiation field slows down convergence of an iterative solution in optically thin parts when transport directions are turned between time steps, since the stored boundary intensities come from the previous solution (see Sect.~\ref{sec:gsmpi} for further details). All ray directions are therefore kept fixed.

The discretized formal solution (Eq.~\ref{eqn:discretefs}) in the simulation domain and averaging of the radiation field over solid angle will be abbreviated in the following using the $\Lambda$ operator, which is commonly defined through
\begin{equation}
J=\Lambda S.
\label{eqn:lambdaop}
\end{equation}
$\Lambda$ is a linear matrix operator on the source functions $S$ which represents the numerical algorithm used to compute the radiation field in the code.


\subsection{The Gauss-Seidel scheme and MPI parallelization}\label{sec:gsmpi}

\begin{figure*}[!htbp]
\centering
\mbox{
\includegraphics[width=8cm]{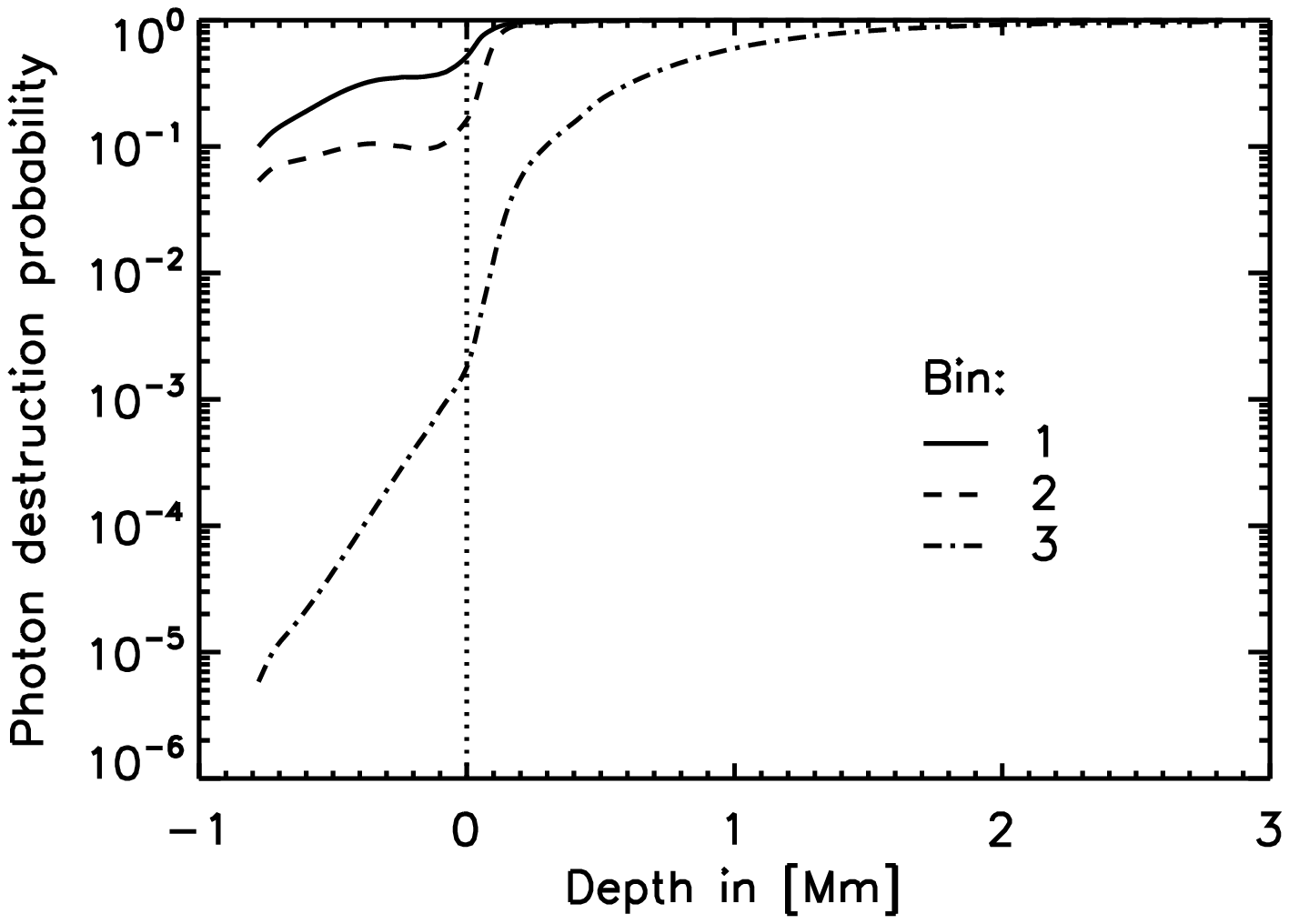} 
\includegraphics[width=8cm]{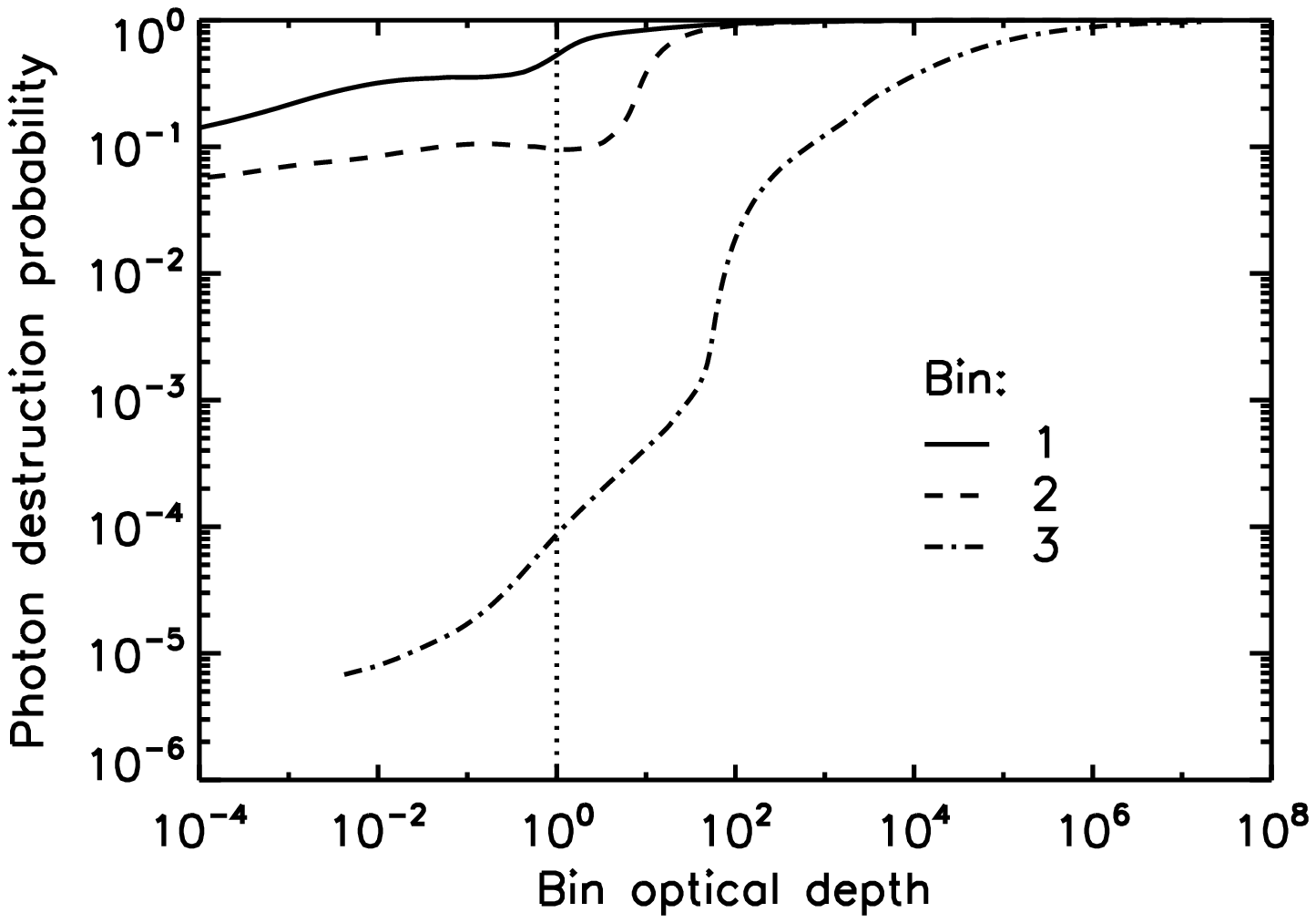} 
}
\caption{Horizontal mean photon destruction probability $\epsilon$ for three bins representing continuum and weak lines (1), intermediate lines (2) and strong lines (3) in the UV, plotted as a function of geometrical depth (left) and optical depth in the respective bin (right, the average is taken over surfaces with the same vertical optical depth). The dotted line marks the stellar surface (left) and unit optical depth in each bin (right).}
\label{fig:epsdepth}
\end{figure*}

As noted in Sect.~\ref{sec:radtrans}, the coherent scattering term turns the transfer equation into an integro-differential equation for the specific intensity $I$. Using the $\Lambda$ operator defined in Eq.~(\ref{eqn:lambdaop}), the problem may be rewritten into the matrix equation
\begin{equation}
\left[\mathbb{1}-(1-\epsilon)\Lambda\right]S=\epsilon B,
\label{eqn:RTmatrixeq}
\end{equation}
with the identity matrix $\mathbb{1}$. The expression represents a very large system of linear equations. Its direct solution in 3D through inversion of the operator on the left-hand side is far too complex and numerically unstable in some cases, to be of practical use. Most solvers therefore apply an iteration scheme, the choice of which depends on the structure of the $\Lambda$ operator matrix. Approximate Lambda Iteration \citep[ALI, sometimes also called Accelerated Lambda Iteration,][]{Cannon:1973} is a popular method to obtain a good approximation of the radiation field with fast convergence. $J$ is computed through a formal solution and used to correct the source function. Rather than just inserting $J$ in $S$, which leads to very slow convergence (or no convergence at all), a largely simplified approximate operator $\Lambda^{\ast}$ is used to compute correction values $\Delta S$ at low cost, speeding up convergence tremendously.

We employ the Gauss-Seidel scheme \citep{TrujilloBuenoetal:1995}, an ALI method that combines the formal solution and correction steps. It mimics a tridiagonal $\Lambda^{\ast}$ operator, but the scheme does not require the expensive construction of the matrix. Source function corrections at the grid point $i$ are obtained during a solver sweep from the expression
\begin{equation}
\Delta S_{i}=\frac{(1-\epsilon_{i})J_{i}^{\mathrm{old/new}}+\epsilon_{i}B_{i}-S_{i}^{\mathrm{old}}}{1-(1-\epsilon_{i})\Lambda_{ii}}.
\end{equation}
$J_{i}^{\mathrm{old/new}}$ is the radiation field that includes the corrections in the upwind part of the simulation domain, which have already been computed during the current sweep. The dependence of $\Delta S_{i}$ on $J_{i}$ in each layer for immediate correction of $S_{i}$ during the sweep requires employing the short characteristics method. The denominator contains the diagonal element $\Lambda_{ii}$ of the $\Lambda$ operator, which may be computed using Eq.~(\ref{eqn:discretefs}) and therefore reduces to a sum of $\Psi$ constants. Source function corrections may be applied during both upsweeps and downsweeps for faster convergence.

We tested our radiative transfer code by comparing the numerical results with an analytical solution for the case of an isothermal 1D atmosphere with constant photon destruction probability $\epsilon$ \citep[see the discussion in][]{TrujilloBuenoetal:1995} and found very good agreement.

The radiation solver is parallelized using spatial domain decomposition and communication with the MPI library, adopting the virtual topology given by the MHD solver of the \texttt{BIFROST} code. The grid is decomposed into cuboid subdomains, allowing an arbitrary number of divisions on all three spatial axes. While this parallelization lends itself to a mixed initial and boundary value problem found in computational hydrodynamics, it is harder to apply in an efficient way to the pure boundary value problem of time-independent radiative transfer. Concurrent computation of spectral subdomains (or opacity bins) would provide a higher degree of parallelism considering the non-local dependencies in a monochromatic formal solution of our given coherent scattering problem, but such an approach would cause severe load balancing issues and suffer from node memory limitations when applying the code to very large simulations. Spatial domain decomposition may still be combined with spectral domain decomposition if radiative transfer needs to be solved for a large number of wavelengths.

\citet{Heinemannetal:2006} have presented a domain-decomposed method based on a variant of the formal solution (Eq.~(\ref{eqn:fs})) on long characteristics. The solver bypasses the problem of missing incident intensities at subdomain boundaries by splitting the local and boundary contributions. While their approach efficiently solves the radiative transfer equation without scattering, the long characteristics solver would have to be combined with a different ALI scheme than Gauss-Seidel. An approximate $\Lambda^{\ast}$ operator needs a certain bandwidth around its matrix diagonal to achieve good convergence \citep[see, e.g., the discussion in][]{Hauschildtetal:2006}. It is therefore more expensive to construct and invert than the diagonal operator used for the Gauss-Seidel scheme.

Our code iterates the solution, starting with the source function and subdomain boundary intensities from the previous hydrodynamical time step, until the maximum relative source function correction in the domain after the $n$th iteration is smaller than a preset threshold $C$:
\begin{equation}
\max\left(\frac{\left|S_{i}^{n}-S_{i}^{n-1}\right|}{S_{i}^{n-1}}\right)\leqslant C.
\end{equation}
When scattering is not included, the maximum relative change of mean intensities at the boundary is used instead to test the convergence of the radiation field:
\begin{equation}
\max\left(\frac{\left|J_{i}^{n}-J_{i}^{n-1}\right|}{J_{i}^{n-1}}\right)\leqslant C.
\end{equation}
If too few iterations are performed, the subdomain boundaries produce artifacts in the upper parts of the atmosphere, where photon mean free paths are comparable to or larger than the subdomain size. In practice, it turns out that a threshold of $C\sim10^{-3}$ yields good results in either case.

\begin{figure*}[!htbp]
\centering
\includegraphics[width=12cm]{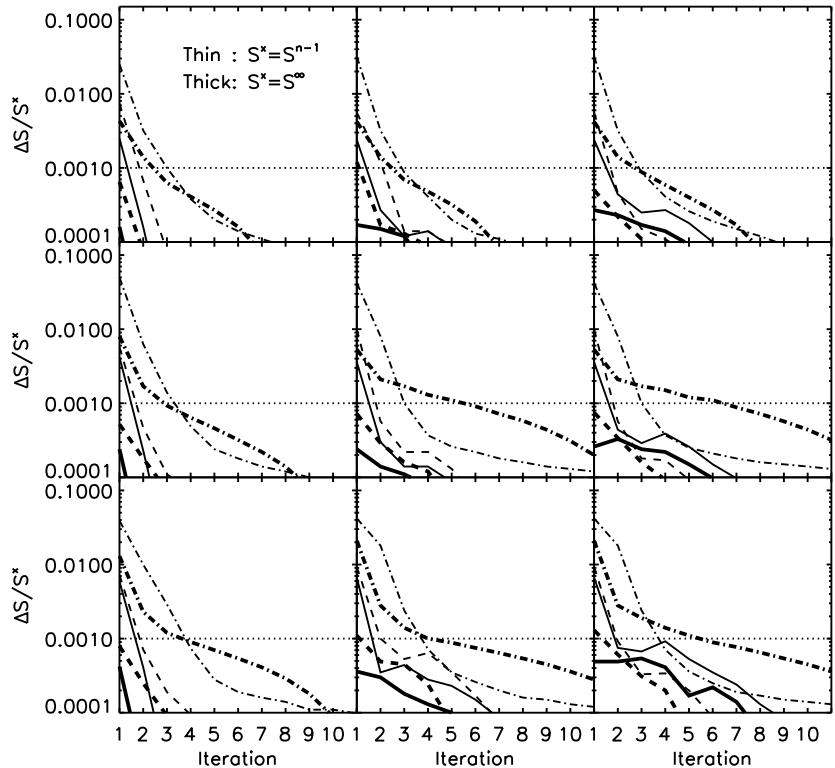}
\caption{
Convergence of the source function for bins 1-3 (see Fig.~\ref{fig:epsdepth}) during a simulation run without domain decomposition (left column), with $2\times2\times2$ decomposition (center column) and $3\times3\times3$ decomposition (right column), and with a time step of $\Delta t=0.03$\,s (upper row), $\Delta t=0.04$\,s (center row) and $\Delta t=0.08$\,s (lower row). Line styles represent the same bins as in Fig.~\ref{fig:epsdepth}. Thin lines: relative source function correction $\Delta S$ after $n$ iterations with respect to $S^{n-1}$ from the previous iteration $n-1$. Thick lines: relative source function correction $\Delta S$ with respect to the ``true'' solution $S^{\infty}$. Dotted lines mark the threshold $C$ beneath which convergence is assumed (see text).
}
\label{fig:Srelabs}
\end{figure*}

The convergence speed of an iterative method depends on the spectral radius $\rho$ of the operator with which corrections are computed, as the error of the solution after $n$ iterations decreases with $\rho^{n}$. The spectral radius approaches $\rho\approx 1-\epsilon$ for optically thick scattering media \citep[see, e.g., the discussion in][]{TrujilloBuenoetal:1995}. Strong scattering at high optical depths therefore leads to very poor convergence rates of the Gauss-Seidel solver, requiring hundreds of iterations in extreme situations. However, this difficulty is mostly alleviated by using the source function solution from the previous time step and the slow evolution of the plasma flow between consecutive time steps, so that the code ideally needs to fully converge the solution only once at the beginning. Domain decomposition additionally slows down convergence if the photon mean free paths cross subdomain boundaries, which is the case at continuum wavelengths in the thin atmosphere, since the subdomain boundary intensities are not initially known. Storing intensities from the previous time step again largely circumvents this problem, and the actual number of iterations per time step that is required during a simulation run depends on how fast the atmosphere evolves.

We therefore test the convergence of the solution for arbitrary time steps of our solar-type simulation using 12 opacity bins with continuum and line scattering (see Sect.~\ref{sec:sun}), following a similar discussion in \citet{Skartlien:2000}. The tests were run at half resolution on all axes to facilitate computation on a single core, which yields slightly faster convergence. Since the true solution $S$ of our radiative transfer problem is unknown, we compare the approximate solution after $n$ iterations, $S^{n}$, with an approximate solution $S^{\infty}$ which we obtained after additional iterations with a lower convergence threshold of $C\sim10^{-4}$, assuming $S^{\infty}\approx S$ with good accuracy.

We use three representative opacity bins, which cover weak, intermediate and strong opacities in the UV, with different depth-dependence of the scattering strengths. The remaining nine bins at longer wavelengths behave in a similar way. Figure~\ref{fig:epsdepth} shows horizontal averages of the photon destruction probabilities $\epsilon$ for each bin in an arbitrary snapshot of our photospheric simulation: averages over layers with the same geometrical depth are plotted in the left panel, averages over surfaces with the same vertical optical depth are plotted in the right panel.

Figure~\ref{fig:Srelabs} compares the convergence speed for the radiative transfer solution of the sample bins with and without domain decomposition, and with different time step lengths. Thick lines represent convergence relative to the true solution $S^{\infty}$ for each bin, thin lines show the convergence relative to the solution obtained in the previous iteration, which we use as the convergence criterion. In normal operation, the solver would stop as soon as the thin line of the currently computed opacity bin has crossed the dotted horizontal line. We caution that the number of iterations needed for a solution also depends mildly on the time stepping algorithm, since the choice of method affects the deviation of stored boundary intensities and source functions between substeps of the time integration. We therefore only analyze the behavior for the first extrapolation step of a 3rd order Runge-Kutta time stepper.

The poorer convergence speed caused by scattering at high optical depths in bin 3 is evident in all plots (thick dot-dashed line), compared to the situation in bin 1, where the photon destruction probability is larger. The small optical path lengths of bin 3 reduce the impact of domain decomposition, since the radiation field is essentially local in most parts of the simulation box. Contrary to that, bin 1 suffers most strongly from slower convergence with increasing number of subdomain divisions, as well as from some flip-flopping of $\Delta S$. The latter is caused by high-order interpolation (see Sect.~\ref{sec:intrefine}) and disappears when the solver is set to linear interpolation. High order interpolation of upwind intensities widens the domain of dependence of the short characteristics, and the effect is amplified where large path lengths in the optically thin regime cross subdomain boundaries.

Domain decomposition mildly slows down convergence, and the accuracy of the solution in bin 3 slightly deteriorates for a larger number of subdomains. Longer time steps have the same effect on that bin, causing slower convergence towards $S^{\infty}$ than indicated by the relative corrections with respect to $S^{n-1}$ (thick and thin dot-dashed lines in Fig.~\ref{fig:Srelabs}). The method devised by \citet{Skartlien:2000} exhibits similar behavior for bins with strong scattering lines.

The effect of such inaccuracies in the numerical solution of $S$ and $J$ on the energy transfer between the radiation field and the gas are nevertheless small or even vanish in some regions: radiative heating is reduced in the atmosphere where coherent scattering is important (see Eq.~(\ref{eqn:divF}) and Fig.~\ref{fig:epsdepth}). Coherent scattering also effectively damps the impact of any remaining discontinuities in the radiation field across subdomain boundaries on the flux divergence in the optically thin atmosphere, so that no visible artifacts from the domain decomposition remain in the gas temperatures.

Compared to the solver proposed by \citet{Heinemannetal:2006}, it is clear that our method is not optimal for the case without scattering, since several computationally expensive formal solution and communication steps are required to obtain a radiation field that is consistent in the whole domain. It offers good performance when scattering is included, which is not considered in their method.
 

\subsection{Interpolation and grid refinement}\label{sec:intrefine}

\begin{figure*}[!t]
\centering
\mbox{
\includegraphics[width=7cm]{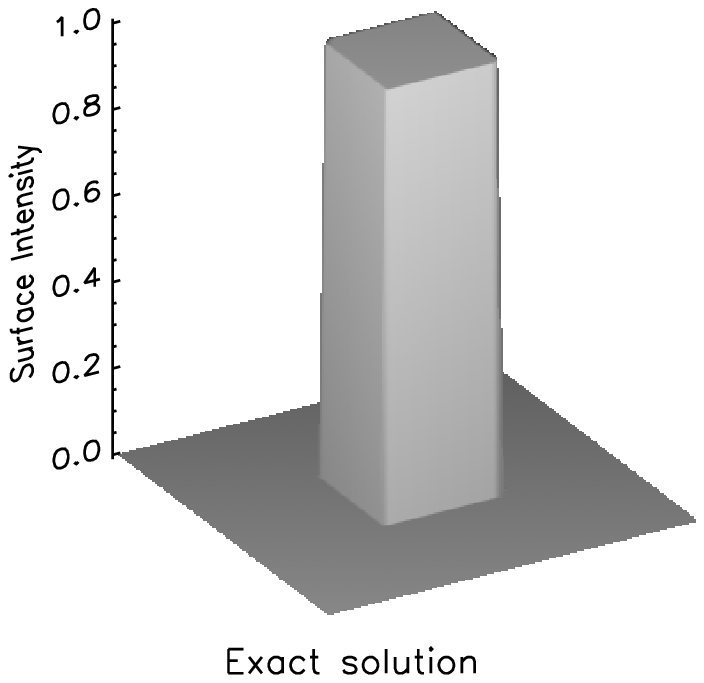}
\includegraphics[width=7cm]{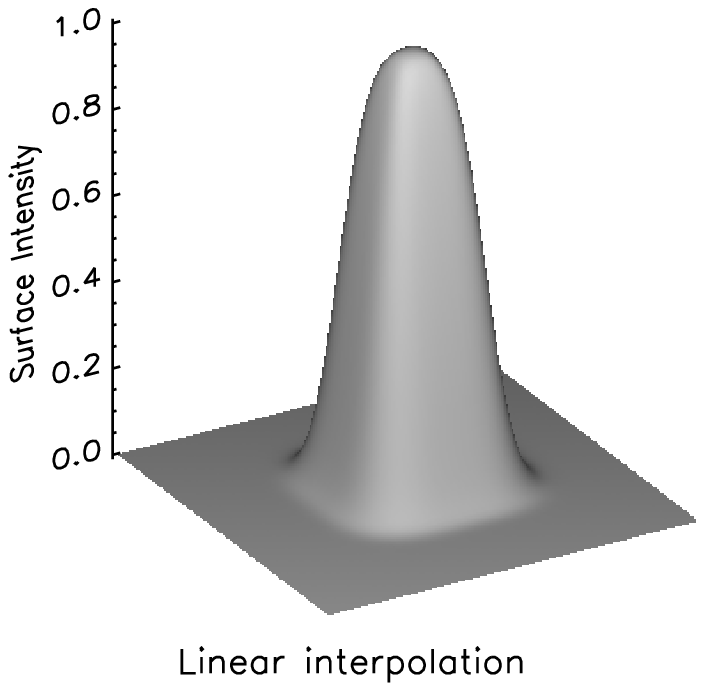}
}
\mbox{
\includegraphics[width=7cm]{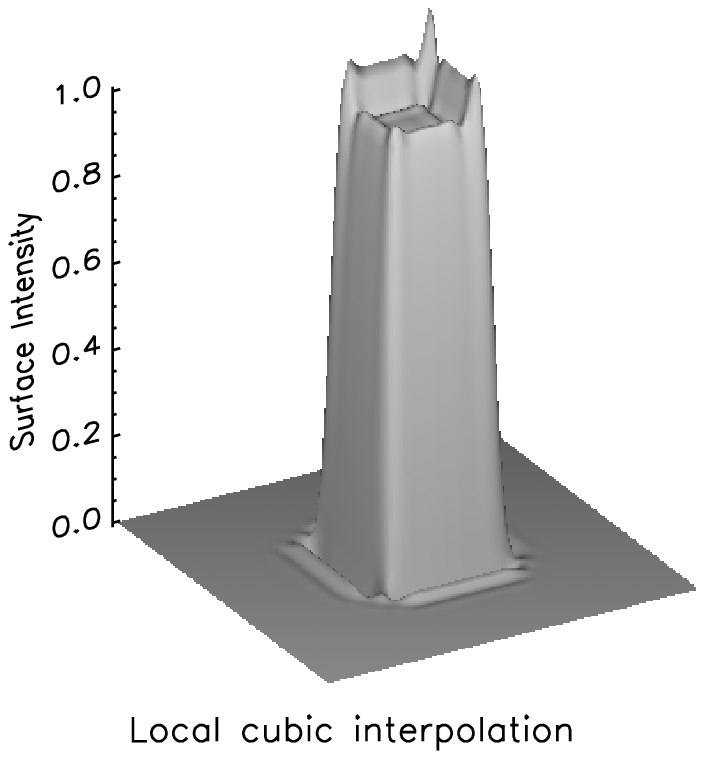}
\includegraphics[width=7cm]{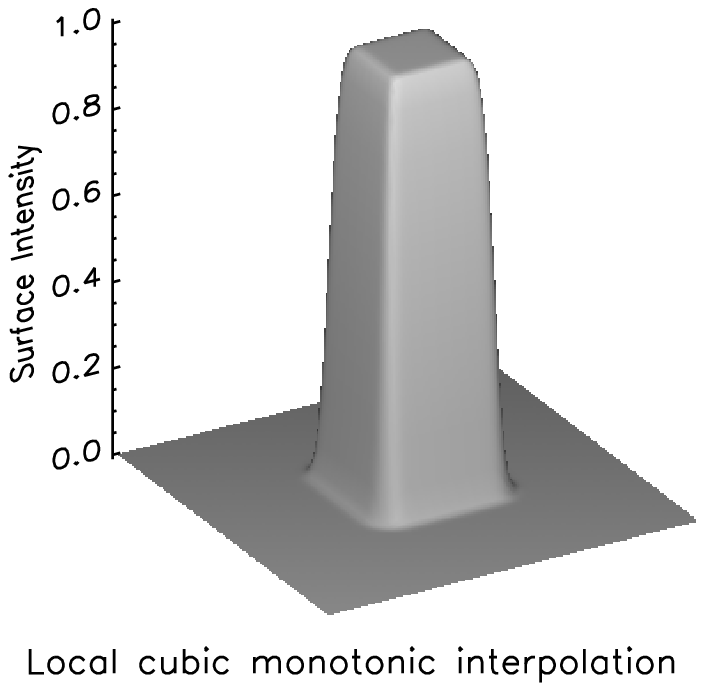}
}
\caption{Numerical diffusion of a searchlight beam with rectangular cross-section using linear (upper right), local cubic (lower left) and local cubic monotonic (lower right) interpolation, compared to the exact solution (upper left).}
\label{fig:searchlight}
\end{figure*}

At every time step, the hydrodynamical solver updates mass densities and internal energy densities. These quantities are used to look up tabulated opacities, bin-integrated Planck functions and photon destruction probabilities at every grid point. In general, the characteristics grid needed to represent the anisotropy of the radiation field does not coincide with the hydrodynamical mesh, requiring the interpolation of $\chi$, $S$ and the upwind intensities $I_{\mathrm{u}}$ during the formal solution.

The accuracy of this interpolation strongly influences the overall accuracy of the solver, and there is a large choice of possible methods \citep[see, e.g., the discussion in][]{Auer:2003}. Linear interpolation is fast and avoids instabilities produced by interpolation overshoots, but yields poor estimates where the radiation field is not well-resolved, e.g. between granules and intergranular lanes at the optical surface. It also amplifies the numerical diffusion effect of short characteristics, where lateral diffusion artificially transports radiation away from the beam.

To illustrate this behavior, we repeat the searchlight test of \citet{Kunaszetal:1988}, where a rectangular light beam is cast through an empty 3D box with a $100^{3}$ mesh and zero opacity. Any diffusion of radiation away from the beam results in a broadening of the beam profile at the surface and can only stem from the interpolation of unattenuated upwind intensities. The light source illuminates the bottom of the 3D box, where it initially covers an area of $30^{2}$ mesh points; it is slanted with an angle of $\theta=28.1^{\circ}$ off the vertical and an azimuth of $\phi=45.0^{\circ}$. The upper left panel in Fig.~\ref{fig:searchlight} shows the beam profile at the top of the 3D box expected from an exact solution of the unattenuated transfer problem through vacuum; note that the finite resolution of the surface in the plot leads to a slightly widened profile. The upper right panel shows the broadening of the beam profile caused by 100 consecutive linear interpolations applied for the numerical transfer through the box. Although the area-integrated intensity is conserved with good accuracy, limited by the machine precision, the beam is visibly widened through numerical diffusion. The lower left panel in Fig.~\ref{fig:searchlight} shows the result when using local cubic interpolation for the transport problem. The broadening is reduced, but the overshooting cubic polynomials produce ringing and negative intensities. We therefore use the local cubic monotonic interpolation scheme of \citet{Fritschetal:1984}, which effectively suppresses overshoots by using weighted harmonic mean derivatives, in consecutive 1D-1D interpolation on horizontal planes, and local quadratic interpolation on vertical cell walls (see Appendix~\ref{sec:cubmonint} for further details). The lower right panel in Fig.~\ref{fig:searchlight} shows the result from the searchlight test, where the beam profile is conserved to a satisfactory degree. Numerical diffusion is reduced and reaches a level which renders the computed flux divergences comparable to those obtained with long characteristics codes: although upwind intensities do not need interpolation along the beam, diffusion affects the local flux divergences when transfered from the slanted long characteristics grid back to the hydrodynamical grid.

The basic mesh on which radiative transfer is computed is imposed by the MHD solver. This is usually not critical in the optically thin upper atmosphere and the optically thick interior, where radiative transfer is simple and may even be over-resolved. The opposite is the case in the transition region around the optical surface, where opacities drop rapidly due to their strong temperature dependence and cause a runaway cooling effect \citep{Steinetal:1998}. For a solar simulation, 1D tests performed by \citet{Nordlundetal:1991} indicate that a vertical spacing of $\lesssim10$\,km is desirable at this atmospheric height. Using a non-linear vertical grid with the finest resolution around the surface, this is easily achievable in 3D for modern MPI-based domain decomposed radiative hydrodynamics codes. However, for large coronal simulations or in the case of giant stars, where the spatial scales needed to resolve hydrodynamics and radiation transport exhibit much larger disparity than in the Sun, finding the optimal grid leads to a conflict. Besides the larger simulation size, too small length intervals $\Delta x$ drastically increase the stiffness of the hydrodynamical equations, where the stability-limited time steps of the transport and diffusion terms scale with $\Delta x$ and $\Delta x^2$, respectively, and quickly become exceedingly small. In extreme cases, both effects may increase computation times of a model beyond tractability.

A fully adaptive mesh for computing radiative transfer would yield optimal results without affecting the stiffness of the equations, but is difficult to realize in a characteristics method. We achieve partial adaptivity by inserting horizontal layers in the hydrodynamical mesh for the radiative transfer computation, reducing optical path lengths without reducing the time steps. The refinement is based on the maximum vertical gradient of the Rosseland mean opacity in each layer and reassessed in regular intervals. While inserting additional layers slows down convergence of the Gauss-Seidel method (see Sect.~\ref{sec:gsmpi}), this is again overcome by storing the source function from the previous time step.


\subsection{Numerical flux divergences}

Having established a method for numerically computing radiative transfer with coherent scattering in a decomposed simulation domain, we now need to obtain flux divergences $\vec{\nabla\cdot F}$, a derivative of the radiation field.

The right-hand side of Eq.~(\ref{eqn:divF}) involves only local quantities that are defined on the cell centers of the hydrodynamical mesh, where $Q_{\mathrm{rad}}$ is eventually needed, and therefore seems a natural choice. The expression $\chi(J-S)$ is numerically stable in the optically thin regime, where round-off errors of a possibly vanishing difference between $J$ and $S$ are attenuated by the exponential outward decrease of the opacity $\chi$. At the same time, $\chi$ amplifies round-off errors of $(J-S)$ beneath the optical surface, where the radiation field thermalizes ($J\approx B$, also in the scattering case since $\epsilon>0$): the flux divergence again vanishes, but the finite machine precision prevents complete cancellation of the terms.

It is possible to stabilize a short characteristics solver in the whole simulation domain by subtracting $S_{0}$ from the discretized formal solution (Eq.~(\ref{eqn:discretefs})), which yields the modified integration constant $\tilde{\Psi}_{0}=\Psi_{0}-1$. Using this equation, one obtains a monochromatic $Q_{\mathrm{rad},\lambda}(\vec{x},\vec{\hat{n}})$ along each ray. We note, however, that this leads to a deviation between the radiative energy $\iint Q_{\mathrm{rad},\lambda}(\vec{x},\vec{\hat{n}})d\Omega dV$ that is emitted by the gas in the simulation volume $V$ per time unit, and the outgoing radiative flux computed from the specific intensities at the surface: the expressions are not equivalent anymore in their discretized form, and numerical errors affect the two values in a different way.

The discretized flux divergence $\vec{\nabla\cdot F}$ on the left-hand side of Eq.~(\ref{eqn:divF}) using finite difference quotients is stable in the optically thick regime, but its accuracy deteriorates outward: round-off errors quickly become significant, as the internal energy per gas volume decreases exponentially \citep[see also the discussion in][]{Brulsetal:1999}.

Adopting the approach presented in \citet{Brulsetal:1999} and \citet{Voegleretal:2005}, we combine both expressions through exponential bridging in each vertical column of the simulation domain as a function of bin optical depth to benefit from their respective advantages. We slightly reduce the transition range between the regimes by a squared exponent, resulting in the expression:
\begin{equation}
Q_{\mathrm{rad}}=e^{-\left(\tau/\tau_{0}\right)^2}Q_{\mathrm{rad}}^J+\left(1-e^{-\left(\tau/\tau_{0}\right)^2}\right)Q_{\mathrm{rad}}^{\vec{F}},
\end{equation}
where $\tau_{0}=0.1$, $Q_{\mathrm{rad}}^J=4\pi\chi(J-S)$ and $Q_{\mathrm{rad}}^F=-\vec{\nabla\cdot F}$, representing the two sides of Eq.~(\ref{eqn:divF}). The total radiative energy computed with this expression delivers a consistent surface flux, since $Q_{\mathrm{rad}}^F$ contributes most of the radiative heating.

Following \citet{Voegleretal:2005}, we compute radiative transfer on cell corners to improve the accuracy of $Q_{\mathrm{rad}}^{\vec{F}}$. Radiative fluxes $\vec{F}$ are averaged over cell corners surrounding each face before computing difference quotients, while $Q_{\mathrm{rad}}^J$ is averaged over all eight cell corners surrounding each grid point. Both expressions use exactly the same stencil and exhibit very good agreement around the threshold optical depth $\tau_{0}$ in our solar-type simulation.

Flux divergences are computed only on the hydrodynamical grid. Additional layers that are possibly inserted by the radiative transfer solver just serve to stabilize the computation and may simply be omitted when computing $Q_{\mathrm{rad}}$, since conservation of the radiative energy flux through the hydrodynamical cell surfaces must hold.

\section{Absorption and scattering opacity sources in the Sun}\label{sec:opacity}

\begin{figure*}[!htbp]
\centering
\includegraphics[width=12cm]{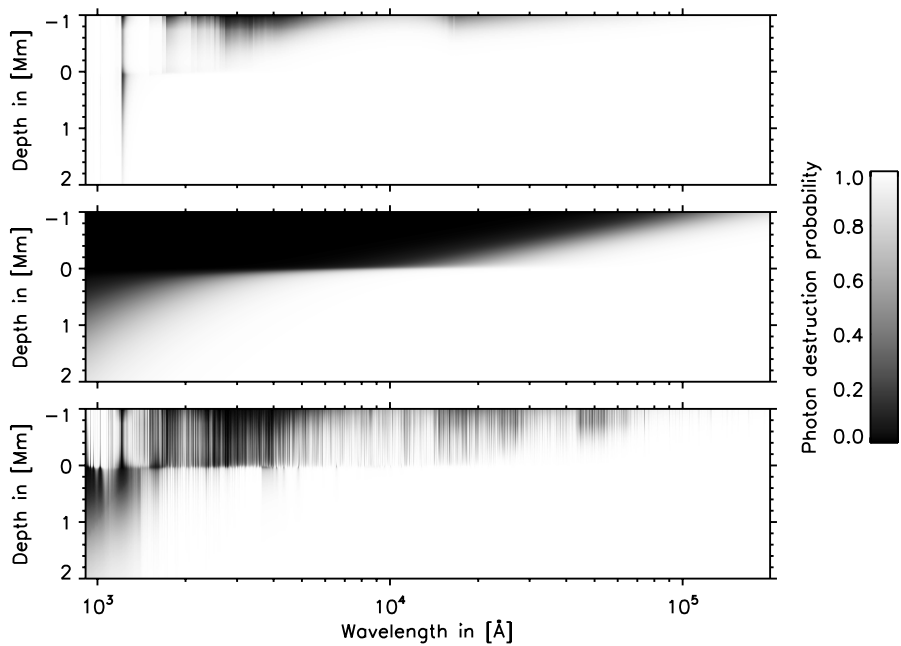} 
\caption{Wavelength and depth dependence of the continuum photon destruction probabilities $\epsilon_{\lambda}^{\mathrm{c}}$ (upper panel), the \citet{Regemorter:1962} line photon destruction probabilities $\epsilon_{\lambda}^{\mathrm{l}}$ (center panel) and the total photon destruction probabilities $\epsilon_{\lambda}$ (lower panel) for the mean solar-type stratification. The zero point on the depth axis marks the stellar surface.}
\label{fig:epssurf}
\end{figure*}

A complete description of radiative transfer in stellar atmospheres requires a detailed wavelength-resolved treatment of numerous radiative absorption and emission processes, collisions with neutral atoms, electrons and ions in the plasma, as well as an evaluation of the feedback of the radiation field on the level populations of the interacting particles. The complexity of the resulting problem vastly exceeds current computational resources. We therefore restrict all of the underlying thermodynamical plasma states to LTE, neglecting the effects of radiation on the excitation and ionization of atoms and photo-dissociation of molecules. The cross-sections and level populations needed for the absorption and scattering coefficients then depend only on the gas density $\rho$ and the temperature $T$. Microscopic plasma thermodynamics is treated with the Mihalas-Hummer-D\"appen equation of state (EOS) for stellar envelopes \citep{Hummeretal:1988,Mihalasetal:1988,Daeppenetal:1988,Mihalasetal:1990} and used in tabulated form. The solar chemical composition for the 15 elements included in the EOS and for the opacities is taken from the abundances of \citet{Asplund:2005}.


\subsection{Continuum opacity}

The most important continuous opacity sources are various transitions of hydrogen atoms, their ions and molecules. The H$^-$ ionization opacity dominates the solar continuum around the optical surface in the visual band; the large temperature sensitivity of the weakly bound second electron in the hydrogen atom causes runaway radiation cooling and the strong temperature gradient found at the top of the granules in the Sun \citep{Steinetal:1998}. Most solar continuum photons originate from this very thin layer. Among many other processes, photoionization of metals contributes significantly to the continuous opacity at shorter wavelengths. Table~\ref{tab:opacities} gives an overview of all sources considered in this work; our data is mostly identical to those used in the latest \texttt{MARCS} models \citep[see Table 1 in][]{Gustafssonetal:2008}, but includes additional bound-free data from the Opacity Project and the Iron Project (see Trampedach et al., in prep., for further details). We also include opacities of the second ionization stage for many metals, allowing 3D models to extend deeper into the convection zone than their 1D counterparts, which is a requirement for correctly simulating surface granulation.

The upper panel in Fig.~\ref{fig:epssurf} shows the wavelength and depth dependence of the continuum photon destruction probabilities $\epsilon_{\lambda}^{\mathrm{c}}$ for the mean stratification of our 3D model, including all continuous absorption and scattering opacity sources considered here. Continuum scattering has a significant contribution mostly above the surface, photons thermalize beneath at almost all wavelengths. Note that the narrow features at the short-wavelength end are the scattering resonances of the Lyman series; Lyman lines are nevertheless treated as true absorbers if line scattering is not included in the simulations. The Rayleigh scattering tail of \ion{H}{I} contributes mostly to the UV continuum opacity in the upper solar photosphere due to its comparatively small cross-section and strong wavelength dependence ($\sigma_{\lambda}\sim\lambda^{-4}$). The importance of elastic scattering on neutral hydrogen is outweighed by thermalizing processes closer to the surface and at short wavelengths. Electron scattering is wavelength-independent in the spectral range considered here, and becomes significant in the upper photosphere, where metals are the most important electron donors. It is mostly notable red-ward of the $1.644$\,$\mu$m edge of H$^-$ bound-free, before H$^{-}$ free-free absorption takes over. Rayleigh scattering on \ion{He}{I} atoms only gives minor contributions to the UV continuum opacity in the upper photosphere. The scattering opacity of H$_{2}$ molecules is negligible. Rayleigh and electron scattering are treated as isotropic, neglecting their weak $(1+\cos^{2}\theta)$ anisotropy, where $\theta$ is the scattering angle away from the incident direction \citep[see, e.g.,][]{Mihalas:1978}.

Between $5000$\,{\AA} and $1.644$\,$\mu$m, the strong H$^{-}$ bound-free absorption opacity thermalizes the photons. Its dominance slightly decreases in the cool outermost layers owing to the lack of free electrons to form the ion.


\subsection{Line opacity}

Spectral line absorption and scattering are important processes which dictate the near-radiative equilibrium found in the solar photosphere. The heating/cooling effect of this line-blanketing forces the flatness of the observed temperature gradient, balancing the adiabatic dynamical gradient; see the discussion in Sect.~\ref{sec:comparison1D3D}. Spectral lines are particularly significant opacity sources at short wavelengths where many radiative bound-bound transitions of metals lie.

We obtain line opacities from extensive opacity sampling tables provided by B. Plez (2008, priv. comm.) as part of the \texttt{MARCS} collaboration. The data are based on VALD with some modifications; see \citet{Gustafssonetal:2008} for further details. The original line data combine scattering and absorption contributions in a total opacity, which is sampled with $\sim100\,000$ wavelengths and tabulated for a range of temperatures and pressures. The tables assume Saha ionization equilibrium and Boltzmann level populations to obtain the absorber density fractions. Departures from LTE, e.g. through radiative ionization, are neglected.

Following \citet{Skartlien:2000}, we estimate the importance of scattering in line transitions by computing a photon destruction probability $\epsilon_{\lambda}^{\mathrm{l}}$ for every line opacity sample, using the \citet{Regemorter:1962} formula (see Appendix~\ref{sec:regemortereps}). We assume all scattering atoms to be neutral, accounting for the large contribution of \ion{Fe}{I} to the line-blanketing \citep{Anderson:1989}, and all transitions to be permitted, in which case the assumptions of the \citet{Regemorter:1962} formula yield reasonable estimates. Only electrons are taken into account for collisional de-excitation. The estimated photon destruction probability $\epsilon_{\lambda}^{\mathrm{l}}$ is then a function of wavelength, temperature and electron pressure, and independent of the actual transition. It may therefore also be applied in cases where the line opacity sample includes several transitions (see the discussion in Appendix~\ref{sec:regemortereps}). Line transitions are treated as independent two-level processes without taking the coupling of the respective level populations into account, which is a reasonable assumption for resonance lines.

The center and lower panels in Fig.~\ref{fig:epssurf} show the wavelength and depth dependence of the estimated $\epsilon_{\lambda}^{\mathrm{l}}$ of spectral lines and the total photon destruction probabilities $\epsilon_{\lambda}$, including all considered continuous and line processes. It is clear that collisional de-excitation dominates beneath the surface and at the longest wavelengths. Resonant line scattering becomes important towards optical and shorter wavelengths at increasing depth.

With the exception of very strong lines, line scattering is generally not coherent due to the Doppler shifts in the moving gas, which are not accounted for in our calculations. The two-level approximation probably gives a reasonably realistic picture of strong permitted lines, but departures from the LTE populations of the atomic levels are still neglected. The important \ion{Fe}{I} opacity deviates from the LTE estimate in higher layers \citep[see Fig. 7 in][]{Shortetal:2005}, thereby affecting the overall magnitude of the line-blanketing in these regions. Moreover, the accuracy of the opacity sampling method itself deteriorates outwards, where fewer and fewer lines contribute to the opacity. The van Regemorter approximation assumes resonant line scattering and consequently produces poorer estimates for all non-resonant lines. In summary, we should expect to obtain an order-of-magnitude estimate for the effects of scattering on the atmospheric structure. A more detailed picture requires a full treatment of the departures from LTE level populations and velocity fields, which is still out of reach for time-dependent 3D simulations.


\section{The effects of scattering on the photospheric temperature structure of a solar-type star}\label{sec:sun}

\subsection{The 3D hydrodynamical surface convection model}\label{sec:hydromodel}

\begin{figure}[!t]
\centering
\includegraphics[width=\linewidth]{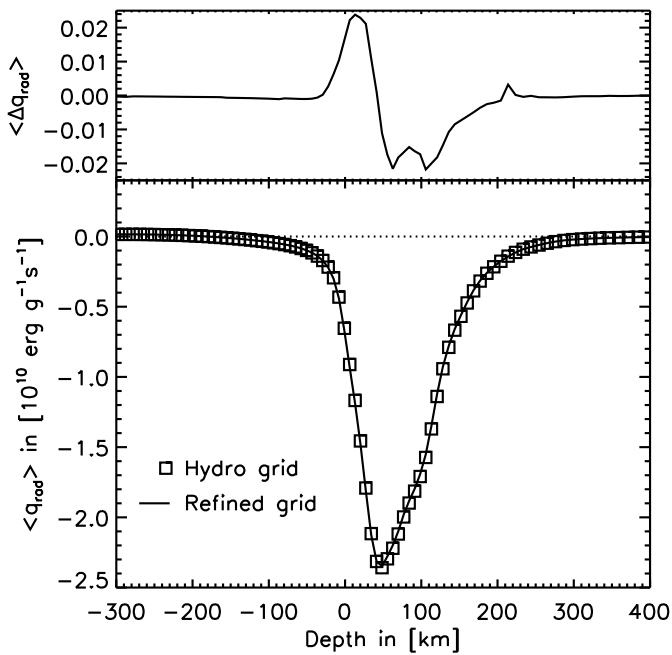}
\caption{Horizontal average heating rate per unit mass around the stellar surface at an arbitrary time step of the simulation. Boxes show $\left<q_{\mathrm{rad}}\right>$ when computed on the hydrodynamical grid; the vertical resolution reaches $7$\,km around the peak. The solid line shows the result after inserting two additional horizontal layers in each hydrodynamical cell. The upper panel gives the average deviation between the two cases in the same units.}
\label{fig:qradsolar}
\end{figure}

To investigate the effects of scattering on the atmosphere of a solar-type star, we conduct time-dependent radiative hydrodynamical simulations of the quiet surface, neglecting the effects of magnetic fields. We solve the fully compressible Navier-Stokes equations, the mass conservation equation and the energy equation, along with the time-independent radiative transfer equation (Eq.~(\ref{eqn:RTtau})); see, e.g., \citet{Steinetal:1998} and \citet{Nordlundetal:2009} for further details. Our $240\times240\times226$ model covers a horizontal area of 6\,Mm $\times$ 6\,Mm at a constant resolution of 25\,km, and extends approximately 700\,km above and 2.8\,Mm below the surface. The vertical resolution reaches 7\,km around the radiative cooling peak and decreases in the optically thick and thin parts of the simulation; radiative transfer is thus resolved well enough that only $\sim3$\,\% of the rays would be affected by overshoots (see Sect.~\ref{sec:SC}). We test the accuracy of the vertical resolution using the adaptive refinement, inserting two extra layers before each computation of radiative transfer. Local differences between the two calculations reach $\sim3\cdot10^{10}$\,erg g$^{-1}$ s$^{-1}$, owing to the strong sensitivity of the heating rate per unit mass, $q_{\mathrm{rad}}\equiv Q_{\mathrm{rad}}/\rho$, to the local temperature gradients in the highly inhomogeneous granulation flow. On the average, however, the change in radiative flux divergence is negligible (see the upper panel of Fig.~\ref{fig:qradsolar}), and the radiation field is well resolved on the hydrodynamical grid. Note the difference between the magnitude of the cooling peaks in Fig.~\ref{fig:qradsolar} and Fig.~\ref{fig:qrad_1D}: the 1D calculation is based on the mean structure; in the 3D case, the average over each depth layer in the 3D box is taken and thus includes lateral inhomogeneities produced by the granulation flow.

Horizontal boundaries are periodic to mimic an infinitely extended atmosphere, vertical boundaries at the top and bottom of the simulation box are open to minimize the interference with the granulation flow. Mass conservation is ensured at the bottom by keeping the gas pressure constant; the underlying convection zone is mimicked by setting the entropy of the inflowing gas. The upper atmosphere is stabilized by setting internal energies to a slowly evolving average at the top.

We approximate the wavelength integral (Eq.~(\ref{eqn:wavint})) with 12 opacity bins to account for the depth-dependence and wavelength-dependence of the absorption and scattering coefficients. The simulation box extends far into the optically thin atmosphere with $\left<\tau_{5000}\right>\approx10^{-6}$, where irradiation $I_{\mathrm{top}}^{-}$ from above is negligible. Rosseland optical depths at the bottom typically reach $\left<\tau_{\mathrm{Ross}}\right>\approx10^{7}$, where radiative transfer is entirely diffusive and the radiation field is completely thermalized. We therefore set the diffusion approximation $I_{\mathrm{bot}}^{+}=B_{\mathrm{bot}}+dB/d\tau$ for all ingoing intensities at the bottom.

The three simulations discussed in Sect.~\ref{sec:tempstruct3D} have mean effective temperatures $T_{\mathrm{eff}}$ between 5804\,K and 5811\,K with average temporal fluctuations of about 13\,K; they are thus slightly hotter than the Sun. For our purposes, there is no need to exactly reproduce the solar $T_{\mathrm{eff}}$. The simulations yield time-series of snapshots spanning $\sim1$\,h of stellar time each, covering several granule lifetimes ($t\sim10$\,min) and several periods of the dominant p-mode ($t\sim5$\,min). Our simulation box covers about 10 granules with typical sizes of the order of $\sim1$\,Mm, allowing us to obtain a statistically meaningful sample of the surface flow in terms of the ergodic hypothesis. The model without scattering was computed with a coarser radiation time step of 0.2\,s, keeping the radiation field constant during the intermediate hydrodynamical calculations. The slow evolution of the flow field and the locality of the Planck source function allow such reduction of the computation time in very good approximation.

\subsection{Scattering in the 1D mean stratification}\label{sec:1Dmean}

\begin{figure*}[t]
\centering
\includegraphics[width=8cm]{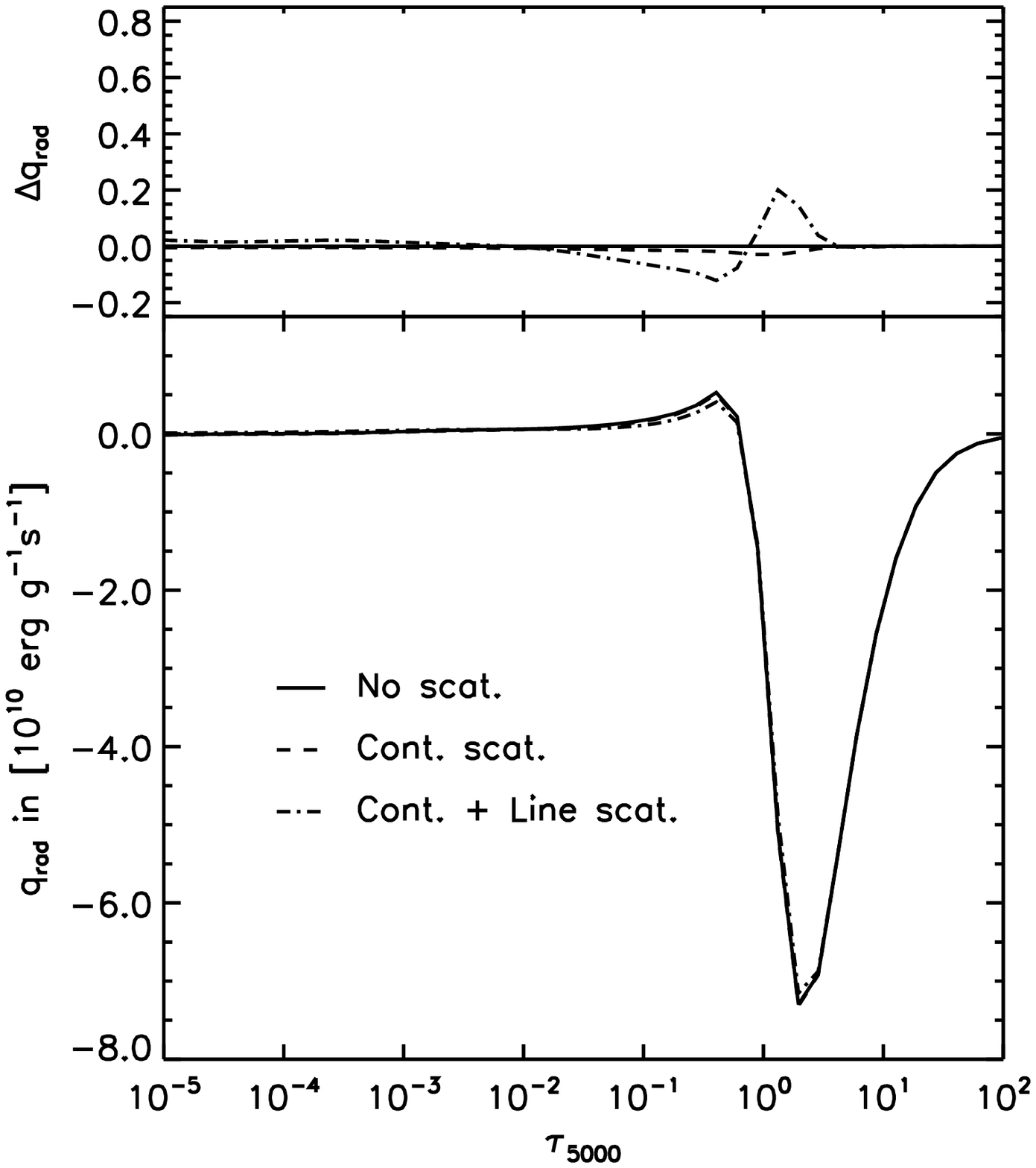}
\includegraphics[width=8cm]{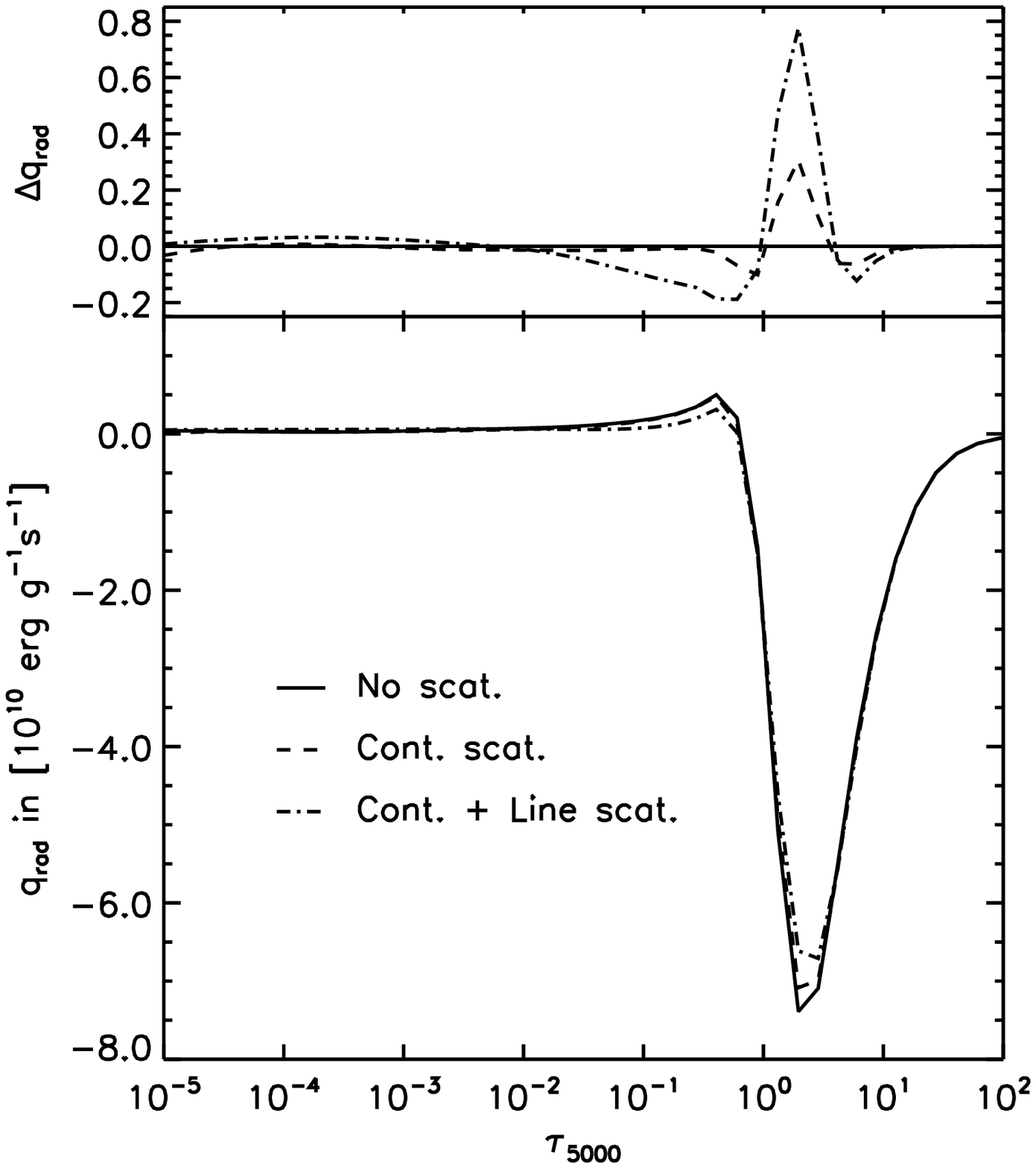}
\caption{\emph{Left}: Heating rates $q_{\mathrm{rad}}$ per unit mass as a function of monochromatic optical depth at 5000\,{\AA}, computed on the 1D mean structure with full opacity sampling for three cases: without scattering (solid line), with continuum scattering (dashed line), and with continuum and line scattering (dot-dashed line). The upper panel shows the deviations of the latter two cases from the computation without scattering (same axis units as in the lower panel). \emph{Right}: Same computation, but using mean opacities and scattering albedos in 12 bins for the radiative transfer computations.}
\label{fig:qrad_1D}
\end{figure*}

We first test the importance of scattering in the 1D mean stratification of our 3D model (the $S=B$ case, see Sect.~\ref{sec:tempstruct3D}) by comparing the wavelength-integrated $q_{\mathrm{rad}}$, using the full opacity-sampled spectrum. Radiative transfer was computed in 1D using a direct block matrix Feautrier-type solver with coherent scattering \citep[for a detailed description see, e.g.,][]{Rutten:2003} and 4th order Radau quadrature for the integral over the polar angle. The left-hand side of Fig.~\ref{fig:qrad_1D} shows $q_{\mathrm{rad}}$ without scattering and $S=B$, with continuum scattering only, and with both continuum and line scattering (lower panel), as well as the deviations from the first case (upper panel).

Continuum scattering seems to have very little impact on $q_{\mathrm{rad}}$ for the given mean structure; the cooling is slightly stronger near the surface. This behavior is expected from the mostly large photon destruction probabilities $\epsilon_{\lambda}^{\mathrm{c}}$ shown in the upper panel of Fig.~\ref{fig:epssurf}.

The differences are slightly larger when scattering is included in the line-blanketing: the small heating bump, where cool uprising gas is heated from beneath by hot granules \citep[see the discussion in ][]{Steinetal:1998}, and the cooling peak beneath the surface both slightly weaken, since the fraction of scattered photons in the line-blanketing does not contribute to heat exchange (cf. the right-hand side of Eq.~(\ref{eqn:divF})). The upper atmosphere, however, now shows slight heating of the mean structure.

We repeat the same test with the binned opacities, computing 1D radiative transfer with and without scattering for 12 mean opacities, photon destruction probabilities and bin-integrated Planck functions. The right panels of Fig.~\ref{fig:qrad_1D} compare again the three different cases. The binning has been optimized for matching sampled and binned $q_{\mathrm{rad}}$ in the $S=B$ case (solid lines in the lower panels of Fig.~\ref{fig:qrad_1D}). The continuum scattering calculation with opacity bins underestimates the cooling beneath the surface. The disparity increases further when line scattering is included; the relative deviations reach 7.5\,\% in the cooling peak (dot-dashed lines in Fig.~\ref{fig:qrad_1D}). However, the overall impact of scattering radiative transfer on the temperature structure of the 3D atmosphere above $\tau_{5000}\gtrsim10^{-3}$ is small (see Sect.~\ref{sec:tempstruct3D} and Fig.~\ref{fig:tttau5}), the same binning setup was therefore adopted for all three simulations. Higher up in the atmosphere, at $\tau_{5000}\lesssim10^{-3}$, opacity binned radiative transfer shows slightly stronger heating of the gas.

\subsection{Scattering in the mean 3D model}\label{sec:tempstruct3D}

\begin{figure}[!t]
\centering
\includegraphics[width=\linewidth]{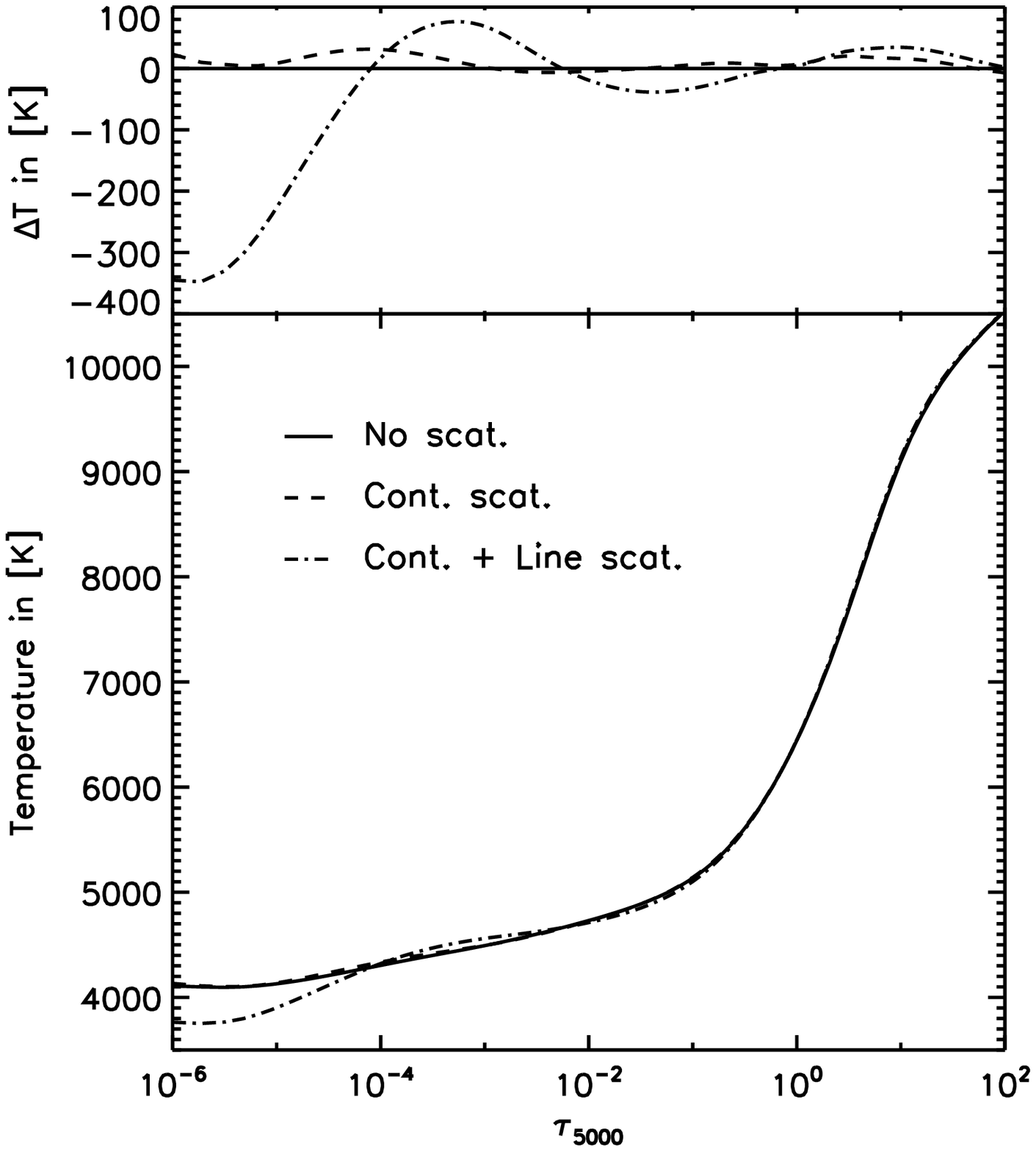}
\caption{Horizontal and temporal average of the mean temperature structure as a function of optical depth at 5000\,{\AA} without scattering (solid line), with continuum scattering (dashed), and with continuum and line scattering (dot-dashed). The upper panel shows the deviation from the first case.}
\label{fig:tttau5}
\end{figure}

In order to assess the effects of continuum and line scattering, we perform three independent simulation runs: the first one treats radiation without scattering by adding all scattering opacity to the absorption opacity and assuming a Planck source function $S=B$. The second one includes continuum scattering in the source function and only adds line scattering opacity to the absorption opacity, and the third one includes scattering both in the continuum and in the line-blanketing. All three time series start from the same initial snapshot and span the exact same amount of simulation time. Snapshots are taken at regular intervals of $\Delta t_{\mathrm{sim}}=10$\,s. We consider time steps at $t_{\mathrm{sim}}>8$\,min after the initial snapshot to allow the atmosphere to adjust to any changes in the radiative heating rates. Exploiting the tight correlation between gas temperature $T$ and vertical optical depth $\tau$ \citep{Steinetal:1998}, we interpolate the 3D temperature cube at each time step of the series onto surfaces with the same optical depth, using a reference $\tau$-scale at 5000\,{\AA}. We then compute the average temperature of each surface in the 3D cube, which yields a 1D mean temperature profile for every snapshot. These profiles are finally averaged over time, and we obtain a very robust characteristic $T$-$\tau$ relation.

Figure~\ref{fig:tttau5} compares the resulting horizontal and temporal mean temperature profiles. The simulations without scattering and with continuum scattering have practically identical stratifications, as expected from the continuum photon destruction probabilities $\epsilon_{\lambda}^{\mathrm{c}}$ (Fig.~\ref{fig:epssurf}) and the 1D test presented in the previous section; continuum scattering is therefore insignificant for the atmospheric stratification in solar-type stars.

The effects of scattering on line-blanketing in and below the photosphere are also rather weak (dot-dashed line in Fig.~\ref{fig:tttau5}). The gas temperatures above $\tau_{5000}\gtrsim10^{-2}$ deviate up to 40\,K from the stratification without scattering, resulting in a slightly steeper temperature gradient around the surface ($\tau_{5000}=1$). Since our adopted binning setup overestimates the deviations for the 1D mean structure (right-hand side of Fig.~\ref{fig:qrad_1D}), the impact of line scattering is probably even smaller at $\tau_{5000}\gtrsim10^{-2}$. The temperature structure in the lower photosphere is thus hardly affected by scattering. The opposite is the case in the high photosphere and above ($\tau_{5000}\lesssim10^{-4}$), where we observe temperatures that are about 350\,K lower, resulting in a significantly steeper mean gradient.

\subsection{Comparison of the 1D and 3D calculations and with other model atmospheres}\label{sec:comparison1D3D}

\begin{figure*}[!htbp]
\centering
\includegraphics[width=12cm]{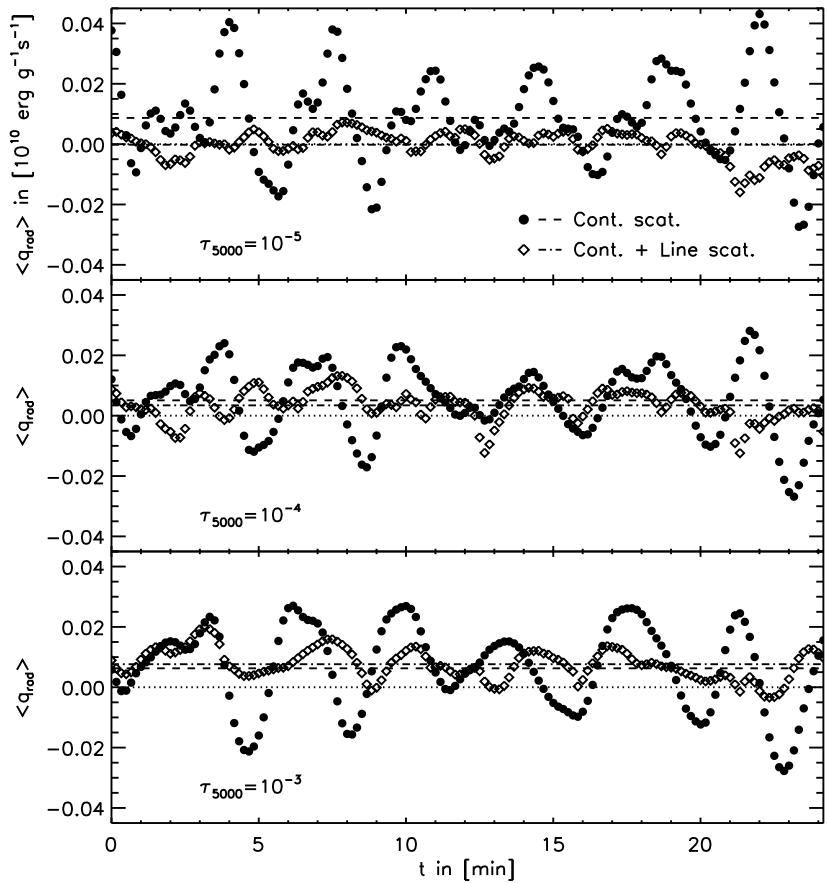}
\caption{Horizontal averages of the radiative heating rates $q_{\mathrm{rad}}$ for the continuum scattering case (circles) and the continuum and line scattering case (diamonds) as a function of simulation time $t$, at optical depths $\tau_{5000}=10^{-5}$ (upper panel), $\tau_{5000}=10^{-4}$ (center panel) and $\tau_{5000}=10^{-3}$ (lower panel). Dashed lines show the spatial and temporal averages for the continuum scattering case, where line scattering is treated as true absorption; dot-dashed lines show the spatial and temporal averages for the line scattering case. Dotted lines indicate zero heating.}
\label{fig:qradseries}
\end{figure*}

The effects of line scattering on the temperature structure of the 3D model seem to be opposite of 1D hydrostatic models in radiative equilibrium, where heating of the highest layers rather than cooling is observed. Indeed, the 1D calculations on the mean 3D atmosphere exhibit slight heating in this region when scattering is included (Fig.~\ref{fig:qrad_1D}). The temperature gradient would therefore become shallower if the 1D calculations were iterated under the assumption of radiative equilibrium \citep[see, e.g., the discussion in][]{Rutten:2003}.

The total radiative flux divergence includes several components: hot radiation from deeper layers at short wavelengths dominates the heating of the gas; the steep outward $dB_{\lambda}/dT$ gradient causes a positive growing $(J-S)$ split. The effect declines in higher layers due to the rapidly decreasing opacity (cf. Eq.~(\ref{eqn:divF})). Strong LTE lines may heat or cool the higher atmosphere (since $J\approx B$ in deeper parts), depending on the spectral region and local temperature gradient, which determine the sign of the $(J-S)$ split. Including coherent scattering in line-blanketing effectively reduces both radiative heating and cooling in high layers through the outwards decreasing $\epsilon^{\mathrm{l}}$ (see Fig.~\ref{fig:epsdepth}). As a consequence, strong resonance lines become unimportant for the temperature structure in high layers, and radiative heating at shorter wavelengths decreases.

\begin{figure*}[!htbp]
\centering
\includegraphics[width=12cm]{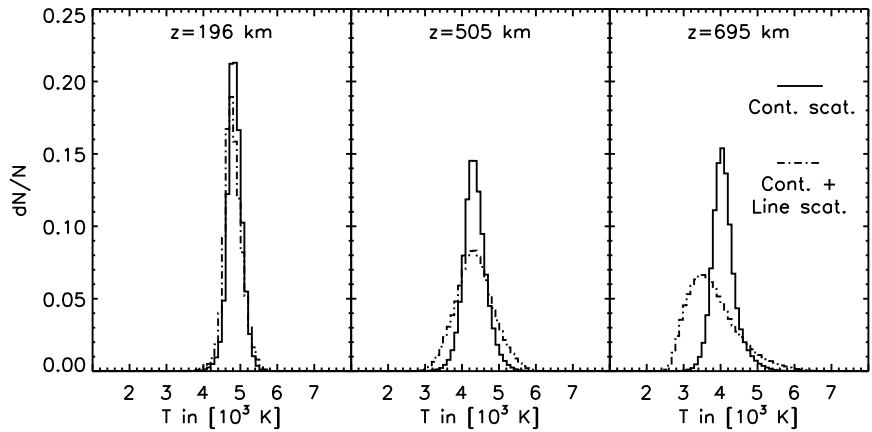}
\caption{Temperature histograms at three different geometrical heights $z$ above the optical surface, integrated over each simulation run. Solid lines show the radiative transfer computation with continuum scattering, dot-dashed lines the case where continuum and line scattering are included. Each temperature distribution is normalized.}
\label{fig:tthist}
\end{figure*}

In the 1D mean atmosphere, scattering-weakened line cooling shifts the total $q_{\mathrm{rad}}$ slightly towards positive values. The behavior of the 3D case can be understood by considering the dynamical nature of our 3D models. Following a derivation in \citet{Mihalasetal:1984}, we insert the continuity equation
\begin{equation}
\frac{D\rho}{Dt}+\rho\vec{\nabla\cdot u}=0,
\end{equation}
where $\rho$ is the gas density, $D/Dt$ is the material derivative and $\vec{u}$ is the gas velocity, into the energy equation,
\begin{equation}
\frac{De}{Dt}+\frac{P}{\rho}\vec{\nabla\cdot u}=q_{\mathrm{rad}},
\end{equation}
where $e$ is the internal energy per unit mass, $P$ is the gas pressure, and $q_{\mathrm{rad}}$ is the radiative heating rate per unit mass; we omit the viscous dissipation term for simplicity. The resulting expression,
\begin{equation}
\frac{De}{Dt}-\frac{P}{\rho^{2}}\frac{D\rho}{Dt}=q_{\mathrm{rad}},
\label{eqn:energyeq}
\end{equation}
is the first law of thermodynamics. An upflowing (downflowing) gas parcel cools (heats) through expansion (compression) represented by the $D\rho/Dt$ term in Eq.~(\ref{eqn:energyeq}), and is exposed to radiative heating through the $q_{\mathrm{rad}}$ term. Equation~(\ref{eqn:energyeq}) is equivalent to the expression
\begin{equation}
T\frac{Ds}{Dt}=q_{\mathrm{rad}},
\end{equation}
where $T$ is the gas temperature and $s$ is the entropy per unit mass, and it is immediately clear that gas motion is adiabatic when $q_{\mathrm{rad}}\rightarrow 0$. In the photosphere of the 3D simulation, temperatures are not affected by scattering. In the upper atmosphere, below $\tau_{5000}\approx10^{-4}$, scattering strongly reduces the line-blanketing. Small or vanishing heating rates $q_{\mathrm{rad}}$ cause the temperature stratification to steepen towards an adiabatic gradient.

Figure~\ref{fig:qradseries} compares the time evolution of radiative heating rates at three optical depths in the atmosphere, averaged over surfaces of constant optical depth to approximately account for vertical gas motion. The plot shows a sequence of snapshots taken at regular simulation time intervals of 10\,s; the zero point on the abscissa is arbitrary. At $\tau_{5000}=10^{-3}$ and $\tau_{5000}=10^{-4}$ (lower and center panels), the continuum scattering case (circles) and the continuum and line scattering case (diamonds) exhibit similar positive heating rates on the average (dashed and dot-dashed lines) and thus similar average temperatures (Fig.~\ref{fig:tttau5}). Line scattering radiative transfer produces slightly stronger mean heating at $\tau_{5000}=10^{-3}$, but fluctuates with lower amplitude. At $\tau_{5000}=10^{-5}$, $q_{\mathrm{rad}}$ practically vanishes on the time average in the line scattering case, but there is still significant radiative heating with line scattering as true absorption. Note the dynamical variation of the sequences: contrary to 1D hydrostatic models, where the radiation field is time-independent by definition, the evolution of the 3D simulations produces fluctuating radiative heating.

\citet{Wedemeyeretal:2004} presented 3D radiation-hydrodynamical simulations of the solar atmosphere that include a chromosphere, using radiative transfer without scattering and solving the equation only for the Rosseland mean opacity to suppress radiative cooling by strong LTE lines. They found an increasing asymmetry of the gas temperature distribution with increasing height above the surface, and a bifurcation in the chromosphere. \citet{Wedemeyeretal:2004} further observed that treating strong spectral lines as true absorption with the opacity binning method reduces the amplitude of temperature fluctuations, which are caused by outward propagating acoustic waves, resulting in unrealistically low maximum temperatures in high layers. \citet{Skartlien:2000} investigated scattering radiative transfer in the chromosphere, comparing radiative heating with and without scattering, and came to the conclusion that including line scattering reduces this damping effect of LTE lines.

Our simulations do not include a chromosphere; the internal energy at the top boundary is set to a slowly evolving mean instead. In the line scattering case, where radiative transfer has only weak influence on the gas, the temperature gradient is sensitive to this boundary condition and thus not well-constrained. However, this does not compromise our conclusions, since the boundary is free to adapt to any upward or downward shift in the mean energies of the gas beneath.

Figure~\ref{fig:tthist} shows temperature distributions of the simulations with continuum scattering and with continuum and line scattering at three different heights above the surface. Our simulations do not reach the same geometrical heights as those of \citet{Skartlien:2000} and \citet{Wedemeyeretal:2004}, and we use a more realistic radiative transfer treatment with 12 opacity bins. We find a similarly growing asymmetry in the temperature distribution of the line scattering simulation in the outer layers \citep[cf. Fig. 7 in][]{Wedemeyeretal:2004}. Treating strong lines as absorbers shifts the mean temperature upward and removes the high temperature tail of the distribution, in qualitative agreement with the findings of \citet{Skartlien:2000} and \citet{Wedemeyeretal:2004}.

\begin{figure}[!htp]
\centering
\includegraphics[width=8cm]{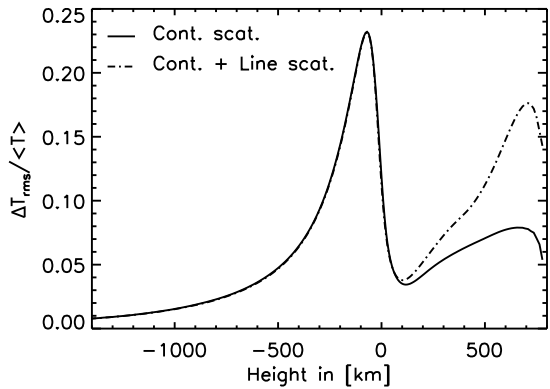}
\caption{Horizontal and temporal averages of the relative temperature fluctuations $\Delta T_{\mathrm{rms}}/\left<T\right>$ as a function of atmospheric height, computed with continuum scattering (solid line) and with continuum and line scattering (dot-dashed line).}
\label{fig:ttrms}
\end{figure}

Figure~\ref{fig:ttrms} shows horizontal and temporal averages of the relative temperature fluctuations, which we define as
\begin{equation}
\frac{\Delta T_{\mathrm{rms}}}{\left<T\right>}=\frac{\sqrt{\left<(T-\left<T\right>)^{2}\right>}}{\left<T\right>}
\end{equation}
in every geometrical depth layer \citep[cf. Eq. 2 and Fig. 9 in][]{Wedemeyeretal:2004}. The comparison between the cases with continuum scattering and with continuum and line scattering confirms the damping of temperature fluctuations through line absorption. Note the decreasing $\Delta T_{\mathrm{rms}}$ at the top of the simulation, which is induced by the hydrodynamical boundary conditions.

We conclude that line scattering is an important ingredient for model atmospheres of solar-type stars that include a chromosphere; while gray radiative transfer reduces damping through strong LTE lines, it cannot produce a realistic temperature structure.

\citet{Anderson:1989} presented simplified, and \citet{Shortetal:2005} presented full 1D non-LTE line-blanketing calculations, respectively, for hydrostatic model atmospheres of solar-type stars. The departures of line-blanketing from LTE through iron-group elements heat up the atmosphere in the height range $10^{-6}\lesssim\tau_{5000}\lesssim10^{-2}$. Our 3D model predicts a predominant temperature decrease as we discussed above. However, it is not clear how departures from LTE in the absorber populations through the ionization balance etc. would affect the atmospheric structure in our 3D simulations, making a direct comparison with the 1D models difficult.

Doppler shifts may have a significant influence on line absorption in higher layers, which expose line cores to hot radiation from deeper in. \citet{Voegleretal:2004} estimated the effects to be insignificant in the photosphere, but their work was based on 1D tests. The large scattering albedo of strong resonance lines, however, should reduce the impact of Doppler shifts higher up.

\section{Conclusions}

We presented a 3D radiative transfer method with coherent scattering for time-dependent (M)HD simulations of stellar atmospheres with the new \texttt{BIFROST} code (Gudiksen et al., in prep.). The simulations are parallelized through domain-decomposition to take advantage of large-scale computer clusters. The solver is based on short characteristics and the Gauss-Seidel scheme for an iterative computation of the radiation field and the radiative flux divergence in the whole simulation domain. We use monotonic interpolation to reduce the numerical diffusion effect of short characteristics and represent the source function integral with B\'ezier polynomials to suppress interpolation overshoots. A partial grid refinement scheme is included to improve the resolution of the radiative transfer computation where strong vertical opacity gradients occur. The wavelength integral is treated in the opacity binning approximation, using 12 bins that divide the opacity spectrum by formation height and wavelength.

The effects of coherent scattering on the temperature structure of a solar-type star are investigated with 3D time-dependent hydrodynamical simulations of magnetically quiet surface convection, including Rayleigh scattering and electron scattering in the continuum and estimated line scattering using the van Regemorter formula. While continuum scattering processes are not important for the mean temperature stratification, we find lower temperatures in the upper atmosphere when scattering is included in the line-blanketing. 3D radiative-hydrodynamical atmospheres thus show the opposite behavior of 1D hydrostatic atmospheres in radiative equilibrium, where scattering in strong lines effectively heats the outer layers.

3D LTE models of solar surface convection have been very successful at reproducing various observational tests, and our results indicate that the solar photosphere is indeed well represented when scattering is not included in radiative transfer. It therefore seems that a refined treatment of the line-blanketing through, e.g., opacity distribution functions or opacity sampling will be the next significant step to improve the realism of 3D radiative-hydrodynamical model atmospheres. Scattering radiative transfer is nevertheless an important ingredient of consistent 3D MHD models of the solar chromosphere, transition region and corona.

While it is not unexpected to see only small differences in the photospheres of solar-type stars when scattering is taken into account, this is likely to change for the much less dense atmospheres of giants, where the importance of Rayleigh scattering increases. The case of metal-poor giants is particularly interesting in that respect, owing to their significance for understanding galactic chemical evolution and the origin of the elements.

\acknowledgements{
This research project has been supported by a Marie Curie Early Stage Research Training Fellowship of the European Community's Sixth Framework Programme under contract number MEST-CT-2005-020395: The USO-SP International School for Solar Physics.
}

\bibliographystyle{aa}
\bibliography{14210}

\begin{appendix}

\section{B\'ezier interpolation of source functions and opacities}\label{sec:scconst}

The discrete formal solution (Eq.~(\ref{eqn:discretefs})) of the radiative transfer equation (Eq.~(\ref{eqn:RTtau})) requires interpolating the source function $S(\tau)$ along the short characteristic. While linear interpolation never overshoots, its accuracy is not sufficient in optically thick media, since the discrete expression is not equivalent to the diffusion approximation. Second-order interpolation significantly improves the accuracy, but suffers from strong overshoots where $\Delta S/\Delta\tau$ gradients change rapidly between the upwind and downwind halves of the characteristic. In extreme cases, this can even destabilize the solver and produce spikes in the local flux divergences.

B\'ezier-type interpolation techniques allow for a direct detection and suppression of such overshoots by virtue of a control point in the polynomial which shapes its curve \citep[see][]{Auer:2003}. A second-order B\'ezier polynomial may be written in the parameterized form
\begin{equation}
S(t)=S_{\mathrm{u}}(1-t)^{2}+S_{0}t^{2}+2S_{\mathrm{c}}t(1-t),
\end{equation}
where $S_{\mathrm{u}}$ and $S_{0}$ are the source functions at the upwind end and the center point of the characteristic, between which interpolation is needed, $t=(\tau-\tau_{\mathrm{u}})/(\tau_{0}-\tau_{\mathrm{u}})$ is the curve parameter, and $S_{\mathrm{c}}$ is the control point. A Bezi\'er curve is always bounded by the convex hull of the three points $S_{\mathrm{u}}$, $S_{\mathrm{c}}$ and $S_{0}$. Using the abbreviations
\begin{equation}
\Delta\tau_{\mathrm{u}}=\tau_{0}-\tau_{\mathrm{u}}; \hspace{0.5cm} \Delta\tau_{\mathrm{d}}=\tau_{\mathrm{d}}-\tau_{0}
\end{equation}
for the optical depths along the characteristic and choosing the control point
\begin{eqnarray*}
S_{\mathrm{c}}	&=& S_{0}-\frac{\Delta\tau_{\mathrm{u}}}{2}S'_{0}\\
			&=& S_{0}-\frac{\Delta\tau_{\mathrm{u}}}{2}\left(\frac{\Delta\tau_{\mathrm{d}}}{\Delta\tau_{\mathrm{u}}+\Delta\tau_{\mathrm{d}}}\frac{S_{0}-S_{\mathrm{u}}}{\Delta\tau_{\mathrm{u}}}+
					\frac{\Delta\tau_{\mathrm{u}}}{\Delta\tau_{\mathrm{u}}+\Delta\tau_{\mathrm{d}}}\frac{S_{\mathrm{d}}-S_{0}}{\Delta\tau_{\mathrm{d}}}\right)\\
\end{eqnarray*}
yields second-order interpolation of $S$, which now also depends on the source function $S_{\mathrm{d}}$ at the downwind end. Introducing the three functions
\begin{eqnarray*}
U_{0}(t) &=& 1-e^{-t} \\
U_{1}(t) &=& t-U_{0}(t) \\
U_{2}(t) &=& t^{2}-2U_{1}(t),
\end{eqnarray*}
and evaluating the integral of the B\'ezier polynomial results in the familiar second-order integration constants for Eq.~(\ref{eqn:discretefs}),
\begin{eqnarray*}
\Psi_{\mathrm{u}} &=& U_{0}(\Delta\tau_{\mathrm{u}})+\frac{U_{2}(\Delta\tau_{\mathrm{u}})-(\Delta\tau_{\mathrm{d}}+2\Delta\tau_{\mathrm{u}})U_{1}(\Delta\tau_{\mathrm{u}})}{\Delta\tau_{\mathrm{u}}(\Delta\tau_{\mathrm{u}}+\Delta\tau_{\mathrm{d}})}\\
\Psi_{0} &=& \frac{(\Delta\tau_{\mathrm{u}}+\Delta\tau_{\mathrm{d}})U_{1}(\Delta\tau_{\mathrm{u}})-U_{2}(\Delta\tau_{\mathrm{u}})}{\Delta\tau_{\mathrm{u}}\Delta\tau_{\mathrm{d}}}\\
\Psi_{\mathrm{d}} &=& \frac{U_{2}(\Delta\tau_{\mathrm{u}})-\Delta\tau_{\mathrm{u}}U_{1}(\Delta\tau_{\mathrm{u}})}{\Delta\tau_{\mathrm{d}}(\Delta\tau_{\mathrm{u}}+\Delta\tau_{\mathrm{d}})}\\
\end{eqnarray*}
\citep[cf. Eq. 8 and 9 in][]{Kunaszetal:1988}. If the source functions $S_{\mathrm{u}}$, $S_{0}$ and $S_{\mathrm{d}}$ have an extremum at $S_{0}$, choosing $S_{\mathrm{c}}=S_{0}$ enforces $S'_{0}=0$, yielding the constants
\begin{eqnarray*}
\Psi_{\mathrm{u}} &=& U_{0}(\Delta\tau_{\mathrm{u}})+\frac{U_{2}(\Delta\tau_{\mathrm{u}})-2\Delta\tau_{\mathrm{u}}U_{1}(\Delta\tau_{\mathrm{u}})}{\Delta\tau_{\mathrm{u}}^{2}}\\
\Psi_{0} &=& \frac{2\Delta\tau_{\mathrm{u}}U_{1}(\Delta\tau_{\mathrm{u}})-U_{2}(\Delta\tau_{\mathrm{u}})}{\Delta\tau_{\mathrm{u}}^2}\\
\Psi_{\mathrm{d}} &=& 0.
\end{eqnarray*}
Overshoots are avoided by limiting $S_{\mathrm{c}}$ to the range of the data points: $\min(S_{\mathrm{u}},S_{0})\leq S_{\mathrm{c}}\leq\max(S_{\mathrm{u}},S_{0})$. If $S_{\mathrm{c}}$ lies outside these limits, choosing $S_{\mathrm{c}}=S_{\mathrm{u}}$ results in the constants
\begin{eqnarray*}
\Psi_{\mathrm{u}} &=& U_{0}(\Delta\tau_{\mathrm{u}})-\frac{U_{2}(\Delta\tau_{\mathrm{u}})}{\Delta\tau_{\mathrm{u}}^{2}}\\
\Psi_{0} &=& \frac{U_{2}(\Delta\tau_{\mathrm{u}})}{\Delta\tau_{\mathrm{u}}^{2}}\\
\Psi_{\mathrm{d}} &=& 0.
\end{eqnarray*}
Note that, contrary to the first two cases, suppressing such overshoots leads to discontinuous left-hand and right-hand derivatives at $S_{0}$.

Optical depths are computed in an analogue fashion to avoid negative results. A second-order B\'ezier polynomial $\chi(s)$ interpolates opacities over the path length $\Delta s$ along the ray; integration over $s$ yields the optical depth interval
\begin{equation}
\Delta\tau=\int_{0}^{\Delta s}\chi(\sigma)d\sigma=\frac{\Delta s}{3}\left(\chi_{\mathrm{u}}+\chi_{0}+\chi_{\mathrm{c}}\right),
\end{equation}
where the control point $\chi_{\mathrm{c}}$ is selected according to the same criteria as discussed above for $S_{\mathrm{c}}$.

\section{Local cubic monotonic interpolation}\label{sec:cubmonint}

The radiative transfer solver uses local cubic interpolation for interpolating data from the hydrodynamical grid onto the characteristics grid. This choice of method improves the accuracy compared to linear interpolation, while ensuring local control of the interpolating polynomial to reduce artifacts. In addition to being a one-pass algorithm, the method also exhibits good computational performance through its instruction-per-data ratio, which is well-suited for modern multi-core CPUs, where high computation speeds are contrasted with slow memory access.

2D interpolation is approximated by consecutive 1D interpolation using a cubic polynomial
\begin{equation}
f(t)=at^{3}+bt^{2}+ct+d,
\end{equation}
with the curve parameter $t\in[0,1]$. The coefficients $a$, $b$, $c$ and $d$ depend on the adjacent data points $f_{1}$ and $f_{2}$ and their derivatives $f'_{1}$ and $f'_{2}$. Inserting the data and reordering the terms, the polynomial may be rewritten in the form
\begin{equation}
f(t)=\alpha(t)f_{1}+\beta(t)f_{2}+\gamma(t)f'_{1}+\delta(t)f'_{2},
\end{equation}
where the interpolation weights $\alpha$, $\beta$, $\gamma$ and $\delta$ now depend on the parameter $t$:
\begin{eqnarray*}
\alpha(t) &=& 2t^{3}-3t^{2}+1\\
\beta(t)&=& 3t^{2}-2t^{3}\\
\gamma(t)&=&\left(t^{3}-2t^{2}+t\right)\Delta x\\
\delta(t)&=&\left(t^{3}-t^{2}\right)\Delta x
\end{eqnarray*}
with the grid spacing $\Delta x$ between the two data points. The shape of the curve is defined by the derivatives $f'_{1}$ and $f'_{2}$. A natural choice is the mean of the left-handed and right-handed difference quotients $f'_{\mathrm{L}}$ and $f'_{\mathrm{R}}$ at both end points. An unweighted arithmetic mean leads to wiggles and overshoots where strong gradients appear. We therefore adopt a recipe by \citet{Fritschetal:1984}, which uses a weighted harmonic mean
\begin{equation}
f'=\left\{
\begin{array}{cc}
\frac{f'_{\mathrm{L}}f'_{\mathrm{R}}}{(1-\alpha)f'_{\mathrm{L}}+\alpha f'_{\mathrm{R}}}   &  \hspace{0.5cm} f'_{\mathrm{L}}f'_{\mathrm{R}} > 0\\
 & \\
0 &  \hspace{0.5cm} f'_{\mathrm{L}}f'_{\mathrm{R}} \le 0
\end{array}
\right.
\end{equation}
with the weighting factor
\begin{equation}
\alpha=\frac{1}{3}\left(1+\frac{\Delta x_{\mathrm{R}}}{\Delta x_{\mathrm{L}}+\Delta x_{\mathrm{R}}}\right),
\end{equation}
which depends on the grid spacing $\Delta x_{\mathrm{L}}$ and $\Delta x_{\mathrm{R}}$ on the left and right sides of the data point. The weighted harmonic mean biases $f'$ towards the smaller of the two difference quotients $f'_{\mathrm{L}}$ and $f'_{\mathrm{R}}$ where strong gradients occur, effectively suppressing overshoots.

Quadratic interpolation uses only one of the two derivatives $f'_{1}$ and $f'_{2}$, depending on the interpolation parameter $t$. The interpolation coefficients are
\begin{equation}
\begin{array}{rclcrcl}
\multicolumn{3}{l}{t\le\frac{1}{2}:} 	&\hspace{0.5cm}	& \multicolumn{3}{l}{t>\frac{1}{2}:}\\ \\
\alpha(t) &=&1-t^2 				&				& \alpha(t)& = & (1-t)^2 \\
\beta(t) &=& t^2 				&				& \beta(t) & = & t(2-t)\\
\gamma(t) &=& t(1-t)\Delta x 		&				& \gamma(t) & = & 0\\
\delta(t) &=& 0 					&				& \delta(t) & = & t(t-1)\Delta x
\end{array}
\end{equation}

\section{Line scattering with the van Regemorter formula}\label{sec:regemortereps}

\begin{figure}[htbp]
\centering
\includegraphics[width=\linewidth]{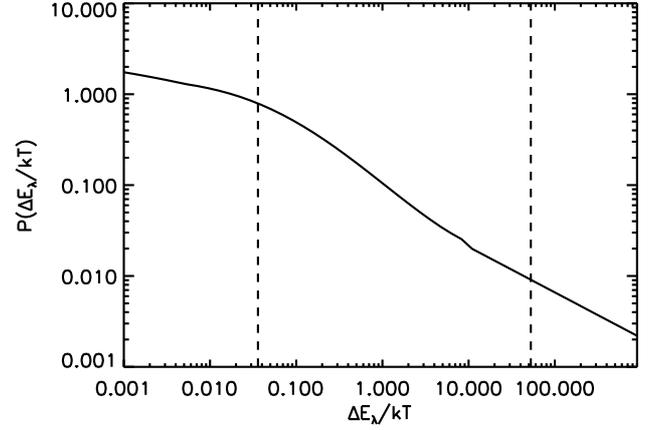}
\caption{Dependence of the tabulated Gaunt factor integral $P(\Delta E_{\lambda},T)$ for collisions of electrons with neutral atoms in the \citet{Regemorter:1962} formula on the transition energy $\Delta E_{\lambda}$ and the gas temperature $T$. The dashed lines mark the boundaries for typical values of $\Delta E_{\lambda}/kT$ found in solar surface convection simulations (Sect.~\ref{sec:hydromodel}).}
\label{fig:regemorterp}
\end{figure}

The photon destruction probabilities in line transitions may be estimated using the semi-empirical \citet{Regemorter:1962} formula to obtain electron collision rates, following the discussion in \citet{Skartlien:2000}. Neglecting other contributions from, e.g., collisions with neutral hydrogen atoms, the de-excitation rate for electron collisions according to this formula is given by
\begin{equation}
C_{21}\sim\lambda^3N_{\mathrm{e}}T^{-1/2}A_{21}P(\Delta E_{\lambda},T),
\label{eqn:elcoll}
\end{equation}
where $\lambda$ is the transition wavelength, $N_{\mathrm{e}}$ is the electron density, $T$ is the gas temperature, and $A_{21}$ is the Einstein coefficient for the corresponding spontaneous radiative transition. The function $P(\Delta E_{\lambda},T)$ abbreviates the velocity integral over the empirically calibrated Gaunt factor of the scattered electron, and depends on the transition energy $\Delta E_{\lambda}$ and the gas temperature $T$. We adopt the tabulated data for neutral atoms of \citet{Regemorter:1962}, see Fig.~\ref{fig:regemorterp}.

The photon destruction probability for a two-level atom is given by
\begin{equation}
\epsilon_{\lambda}=\frac{\kappa_{\lambda}}{\kappa_{\lambda}+\sigma_{\lambda}}=\frac{C_{21}}{C_{21}+A_{21}+B_{21}B_{\lambda}},
\label{eqn:lineeps}
\end{equation}
where $B_{21}$ is the rate for induced de-excitation, and $B_{\lambda}$ is the Planck function. Neglecting the induced de-excitation term, Eq.~(\ref{eqn:lineeps}) simplifies to
\begin{equation}
\epsilon_{\lambda}\approx\frac{1}{1+A_{21}/C_{21}}.
\label{eqn:lineepssimple}
\end{equation}
$\epsilon_{\lambda}$ is independent of the actual transition after inserting the van Regemorter formula Eq.~(\ref{eqn:elcoll}), and thus only a function of $\lambda$, $N_{\mathrm{e}}$ and $T$.

Line opacities in stellar spectra often combine contributions from many transitions at a given wavelength. The total monochromatic photon destruction probability of an opacity sample at wavelength $\lambda$ is given by the sum over all transitions,
\begin{equation}
\epsilon^{\mathrm{l}}_{\lambda}=\frac{\sum_{i}\kappa^{\mathrm{l}}_{\lambda,i}}{\sum_{i}\chi^{\mathrm{l}}_{\lambda,i}}.
\end{equation}
Inserting Eq.~(\ref{eqn:lineepssimple}), thus assuming the above mentioned approximations, yields
\begin{equation}
\epsilon^{\mathrm{l}}_{\lambda}\approx\frac{\sum_{i}\epsilon_{\lambda}\chi^{\mathrm{l}}_{\lambda,i}}{\sum_{i}\chi^{\mathrm{l}}_{\lambda,i}}=\epsilon_{\lambda},
\end{equation}
where the equality holds since $\epsilon_{\lambda}$ is independent of the actual transition $i$. The absorption and scattering contributions $\kappa^{\mathrm{l}}_{\lambda}$ and $\sigma^{\mathrm{l}}_{\lambda}$ to each opacity sample $\chi^{\mathrm{l}}_{\lambda}$ are then isolated using $\epsilon^{\mathrm{l}}_{\lambda}$ and added to the coefficients of the continuum processes (see Table~\ref{tab:opacities}).

\section{Continuum opacity sources}

\begin{table}[htdp]
\caption{Continuum opacity sources}
\begin{tabular}{ll}
\hline\hline
Absorber and process				& Reference		                  \\
\hline
H$^-$ b-f							& \citet{Broadetal:1976,Wishart:1979}		\\
H$^-$ f-f							& \citet{Belletal:1987}		\\
\ion{H}{I} b-f, f-f						& \citet{Karzasetal:1961}		\\ 
\ion{H}{I}+\ion{H}{I}					& \citet{Doyle:1968}			\\ 
\ion{H}{I}+\ion{He}{I}					& \citet{Gustafssonetal:2001}	\\ 
H$_2$+\ion{H}{I}					& \citet{Gustafssonetal:2003}	\\ 
H$_2$+\ion{He}{I}					& \citet{Joergensenetal:2000}	\\ 
H$_2$+H$_2$						& \citet{Borysowetal:2001}		\\ 
H$_2^-$ f-f						& \citet{Bell:1980}				\\ 
H$_2$ photo-dissociation				& \citet{Allisonetal:1969}	\\
H$_2^+$ b-f, f-f						& \citet{Stancil:1994}				\\ 
He$^-$ f-f							& \citet{John:1995}                                 \\ 
\ion{He}{I} b-f						& TOPbase$^{1}$                         \\ 
\ion{He}{I} f-f						& \citet{Peach:1970}			\\ 
\ion{He}{II} b-f						& TOPbase$^{1}$                         \\
C$^-$ f-f							& \citet{Belletal:1988}			\\ 
\ion{C}{I} b-f						& \citet{Naharetal:1991}             \\
\ion{C}{I} f-f						& \citet{Peach:1970}			\\ 
\ion{C}{II} b-f						& \citet{Nahar:1995,Nahar:2002}            \\
\ion{C}{II} f-f						& \citet{Peach:1970}			\\ 
\ion{C}{III} b-f						& \citet{Naharetal:1997}             \\
N$^-$ f-f							& \citet{Ramsbottometal:1992}		\\ 
\ion{N}{I} b-f						& \citet{Naharetal:1997}                               \\ 
\ion{N}{II} b-f						& \citet{Naharetal:1991}                                \\ 
\ion{N}{III} b-f						& \citet{Naharetal:1997}                                \\ 
O$^-$ f-f							& \citet{John:1975}                                 \\ 
\ion{O}{I}, \ion{O}{II} b-f				& \citet{Nahar:1998}                                \\ 
\ion{O}{III} b-f						& \citet{Naharetal:1994b}                                \\ 
\ion{Ne}{I}, \ion{Ne}{II}, \ion{Ne}{III} b-f	& TOPbase$^{1}$                          \\
\ion{Na}{I}, \ion{Na}{II}, \ion{Na}{III} b-f	& TOPbase$^{1}$                          \\
\ion{Mg}{I}, \ion{Mg}{II}, \ion{Mg}{III} b-f	& TOPbase$^{1}$                          \\ 
\ion{Al}{I}, \ion{Al}{II}, \ion{Al}{III} b-f		& TOPbase$^{1}$                          \\ 
\ion{Si}{I} b-f						& \citet{Naharetal:1993}		\\
\ion{Si}{II} b-f						& \citet{Nahar:1995}				 \\ 
\ion{Si}{III} b-f						& TOPbase$^{1}$			 \\
\ion{S}{I} b-f						& TOPbase$^{1}$			 \\
\ion{S}{II} b-f						& \citet{Nahar:1995}                                 \\
\ion{S}{III} b-f						& \citet{Nahar:2000}                                 \\
\ion{Ar}{I}, \ion{Ar}{II}, \ion{Ar}{III} b-f		& TOPbase$^{1}$                           \\
\ion{Ca}{I}, \ion{Ca}{II}, \ion{Ca}{III} b-f	& TOPbase$^{1}$                           \\ 
\ion{Fe}{I} b-f						& \citet{Bautista:1997}                                  \\ 
\ion{Fe}{II} b-f						& \citet{Naharetal:1994a}                                  \\ 
\ion{Fe}{III} b-f						& \citet{Nahar:1996}                                  \\ 
\ion{Ni}{II} b-f					         & \citet{Bautista:1999}                                  \\
CO$^-$ f-f							& \citet{John:1975}				\\ 
H$_2$O$^-$ f-f						& \citet{John:1975}				\\ 
OH b-f							& \citet{Kuruczetal:1987}				\\ 
CH b-f							& \citet{Kuruczetal:1987}				\\ 
\hline
\ion{H}{I} scattering					& \citet{Gavrila:1967}                   \\ 
H$_2$ scattering					& \citet{Victoretal:1969}               \\ 
e$^-$ scattering					& Thomson			           \\ 
\ion{He}{I} scattering					& \citet{Langhoffetal:1974}           \\ 
\hline
\end{tabular}
\label{tab:opacities}

\begin{list}{}{}
\item[$^{1}$] Contains Opacity Project data \citep{Seatonetal:1994}
\end{list}

\end{table}%

\end{appendix}

\end{document}